\documentclass[]{article}

\usepackage[utf8]{inputenc}
\usepackage{amsmath,amsfonts}
\usepackage{amssymb,commath}
\usepackage{authblk}
\usepackage{tikz}
\usepackage{graphicx}
\usepackage[footnotesize,hang]{caption}
\usepackage{subfig}
\usepackage{color}
\usepackage{rotating}
\usepackage{mathtools}
\usepackage{xcolor}
\usepackage{epstopdf}
\usepackage{cite}
\usepackage{url}
\usepackage{pifont}

\usepackage[margin=0.65in]{geometry}
\def\Integer{\mathbb{Z}}

\usetikzlibrary{decorations.pathmorphing}

\renewcommand{\vec}[1]{\boldsymbol{#1}}

\newcommand{\e}{\mathrm{e}}

\newcommand{\diff}[2]{\frac{\mathrm{d} #1}{\mathrm{d} #2}}

\def\XXint#1#2#3{{\setbox0=\hbox{$#1{#2#3}{\int}$}
\vcenter{\hbox{$#2#3$}}\kern-.5\wd0}}

\title{Exponential asymptotics of woodpile chain nanoptera using numerical analytic continuation}
\author[1]{Guo Deng\footnote{Electronic address: guo.deng@mq.edu.au}}
\author[1]{Christopher J. Lustri\footnote{Corresponding Author. Electronic address: christopher.lustri@mq.edu.au}}
\affil[1]{Department of Mathematics and Statistics, 12 Wally's Walk, Macquarie University, New South Wales 2109, Australia}
\date{}

\begin{document}
\maketitle
\abstract{Travelling waves in woodpile chains are typically nanoptera, which are composed of a central solitary wave and exponentially small oscillations. These oscillations have been studied using exponential asymptotic methods, which typically require an explicit form for the leading-order behaviour. For many nonlinear systems, such as granular woodpile chains, it is not possible to calculate the leading-order solution explicitly. We show that accurate asymptotic approximations can be obtained using numerical approximation in place of the exact leading-order behaviour. We calculate the oscillation behaviour for Toda woodpile chains, and compare the results to exponential asymptotics based on tanh-fitting, Pad\'{e} approximants, and the adaptive Antoulas-Anderson (AAA) method. The AAA method is shown to produce the most accurate predictions of the amplitude of the oscillations and the mass ratios for which the oscillations vanish. This method is then applied to study granular woodpile chains, including chains with Hertzian interactions -- this method is able to calculate behaviour that could not be accurately approximated in previous studies.}

\section{Introduction}
\label{intro}

\subsection{Exponential asymptotics}
Exponential asymptotics are techniques used to study behaviour in exponentially small behaviour in singularly-perturbed systems. Important behaviour on exponentially small scales arises in models from quantum mechanics~\cite{Maslov,Joye,Meyer1991_1,Giller2000,Hagedorn},
optics~\cite{wood1991,Chapman2009,Lawless,Meyer1991,Nixon}, crystal growth~\cite{Brower,Kruskal},
fluid mechanics~\cite{Chapman2002,Lustri2013,Lustri2014,Chapman2006,Lustri2020}, particle chains~\cite{Lustri,Lustri1,Deng2021,Deng2022} and cosmology~\cite{Andersson}. 

Classic asymptotic power series cannot describe these exponentially small terms, as they are smaller than any algebraic power of the asymptotic parameter. Exponential asymptotic methods, first developed in~\cite{Berry1988,Berry}, make use of the idea that truncating an asymptotic power series optimally produces an exponentially small remainder. This remainder term can be studied in order to calculate the asymptotic behaviour of the solution on exponentially small scales. The method used in the paper, developed in~\cite{Chapman,Daalhuis}, are described in more detail in Section \ref{s:exponential}.

Exponentially small asymptotic contributions in singularly-perturbed problems typically exhibit behaviour known as Stokes' phenomenon. This describes rapid changes in the exponentially small contribution that occur as curves in the complex plane, known as Stokes curves, are crossed. 

Exponential asymptotic techniques typically require the explicit calculation of the leading-order solution in the asymptotic limit. Stokes curves originate at singularities in the leading-order solution, and this solution also affects the form of the exponentially-small asymptotic behaviour that is switched as these curves are crossed. For many nonlinear problems, such as~\cite{Lustri1,Deng2021,Deng2022,ChapmanTrinh}, it is impossible to calculate the leading-order behaviour explicitly. The leading-order behaviour can be calculated numerically, but this numerical solution must be analytically continued to determine the location of singular points in the complex plane.

An example of a physical system for which the leading-order solution cannot be calculated exactly is the granular woodpile chain studied in~\cite{Deng2022}. A tanh-fitting method based on the work of~\cite{sen2001} was used to approximate the leading-order behaviour. This method produced accurate results for large values of the interaction exponent but failed to capture important behaviour for smaller values, including the important case of Hertzian particle interactions.

In this paper we investigate how numerical approximations to the leading-order behaviour affect the accuracy of using exponential asymptotic methods to study woodpile chains. We perform an exponential asymptotic analysis on nonlinear waves in a Toda woodpile lattice, which has an exact leading-order solution. We then compare three possible approximation methods for determining the leading-order behaviour in the complex plane against the results of the exact analysis: tanh-fitting~\cite{sen2001}, Pad\'{e} approximants~\cite{pade}, and the adaptive Antoulas-Anderson (AAA) method~\cite{Nakatsukasa}. 

We then apply the most effective of these methods (AAA approximation) to produce descriptions of exponentially small behaviour in granular woodpile chains, including the Hertzian woodpile chains that could not be accurately calculated in~\cite{Deng2022}. We show that this inaccuracy occured because tanh-fitting cannot accurately identify the location of subdominant singularities. This problem is avoided using a method based on AAA approximation for the leading-order behaviour, which is able to accurately identify subdominant singularity locations and describe their interaction effects.

\subsection{Woodpile chains}

The motion of homogeneous particle chains, where every particle is identical, is governed by the system of differential--difference equations
\begin{equation}
	m\ddot{x}(n,t)=\phi'(x(n+1,t)-x(n,t))-\phi'(x(n,t)-x(n-1,t))\,,
\label{e:lattice}
\end{equation}
where $n\in\Integer$, dots denote temporal derivatives, $\phi$ is the interaction potential and $\phi'$ is the derivative of $\phi$. It is known that solitary wave solutions exist in lattices with superquadratic potentials~\cite{FrieseckeWattis,Stefanov}. Among lattices with superquadratic potentials, the Toda lattice~\cite{Toda,Toda1,Toda2,Toda3,Toda4} and the granular chain~\cite{Nesterenko,Nesterenko1,sen:2008} have been of particular interest due to their theoretical and practical importance.

The Toda lattice describes particle interactions with interaction potential
\begin{equation}
\phi(r)=\frac{a}{b}(\e^{-br}-1)+ar,
\label{e:potentialToda}
\end{equation}
where $a$ and $b$ are constants. The Toda lattice is a completely integrable Hamiltonian system, allowing for the use of analytical tools to obtain and study explicit solutions for the chain behaviour~\cite{Flashchka1,Flashchka2,Henon,Venakides,Deift}.

\begin{figure}[tb]
\centering
\subfloat[Precompressed particle chain]{
\includegraphics[height=0.23\textwidth]{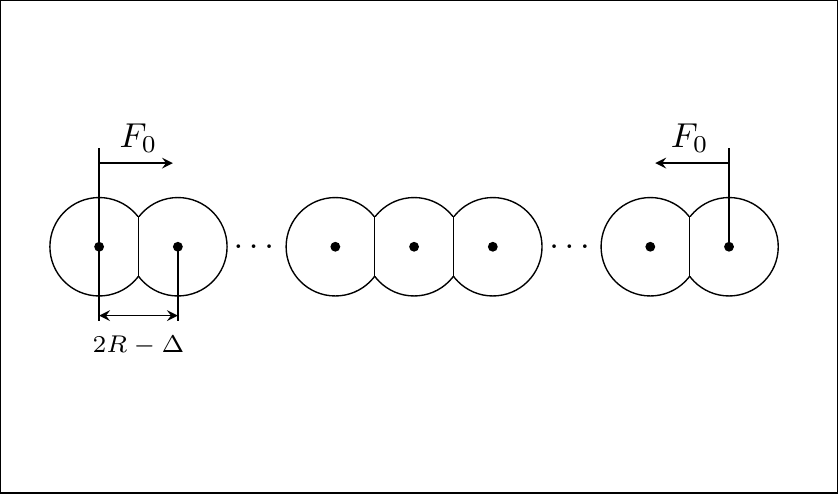}
}
\subfloat[Physical woodpile]{
\includegraphics[height=0.23\textwidth]{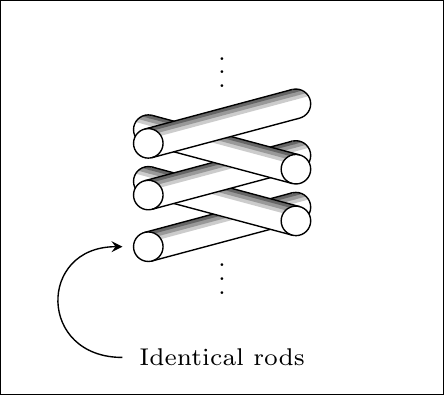}
}
\subfloat[Idealized model]{
\includegraphics[height=0.23\textwidth]{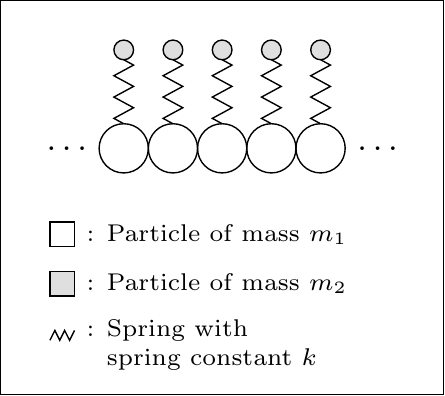}
}
\caption{(a) Schematic illustration of a granular chain with precompression. The chain consists of identical, aligned spheres with radius $R$, with a compressive force $F_0$ applied at both ends of the chain to compress the equilibrium particle positions. (b) The physical configuration of orthogonally stacked rods known as a ``woodpile chain''. (c) An idealized mathematical model of the physical configuration in (b). In this model,
the interaction between heavy particles is given by~\eqref{e:potentialToda} or~\eqref{e:potentialHertz},
and light spherical particles (sometimes called ``resonators'') are attached to each heavy particle by a spring. 
}
\label{f:DiatomicWoodpile}
\end{figure}

Granular chains describe the interaction of physical beads, and the interaction potential is given by
\begin{eqnarray}
	\phi(r)=
\begin{cases}
	c(\Delta-r)^{\alpha + 1}\,, &r\leq\Delta \cr 0\,, &r>\Delta \end{cases}\,,~\quad c = \mathrm{constant}\,,
\label{e:potentialHertz}
\end{eqnarray}
where $\alpha > 1$, $c$ is constant, and $\Delta$ is the equilibrium overlap of adjacent particles due to precompression. Figure~\ref{f:DiatomicWoodpile}(a) illustrates a granular chain with precompression. The interaction potential~\eqref{e:potentialHertz} is only nonzero between particles that are in physical contact, and it is zero for neighbouring particles that are not in contact.

 The granular chain is important from a practical point of view, as it arises in various applications. Previous theoretical, numerical, and experimental studies have explored the generation~\cite{Deng1,Lazaridi,Coste,DaraioNesterenko2006,Hinch}, propagation~\cite{Deng1,Lazaridi,Coste,DaraioNesterenko2006,Hinch}, interaction~\cite{manciu:2002,job:2005,avalos:2009,avalos:2011,avalos:2014,Deng}, and long-time dynamics \cite{avalos:2011,avalos:2014,sen:2004,Prz:2015,przedborski:2017} of solitary waves in granular chains.

A woodpile chain is composed of orthogonally stacked slender rigid cylinders, illustrated in Figure~\ref{f:DiatomicWoodpile}(b). The elastic deformation in the direction perpendicular to the stack direction is modelled by internal resonators, where the mass and coupling constant of these resonators are determined by system properties, such as the mass, shape, and material of the cylinders. For a stack of identical cylinders, each resonator has the identical mass and coupling constant. The stack of identical cylinders can be modelled as a homogeneous chain with a prescribed interaction potential, where each chain particle is also connected to an external particle by a linear spring. We denote the mass of particles in the monatomic chain by $m_1$, the mass of external particles by $m_2$, the mass ratio by $\eta^2=m_2/m_1$ and the spring constant by $k$.

We consider woodpile chains with interaction potential along the direction of the stack given by Toda~\eqref{e:potentialToda} and power law~\eqref{e:potentialHertz} interactions. The configuration of this idealized model of woodpile chains is shown in Figure~\ref{f:DiatomicWoodpile}(c). For the remainder of this study, we use the term ``woodpile chain'' to describe this idealized model.

The governing equations of an idealized woodpile chain are
\begin{align}\label{1:gov0a}\nonumber
	m_1 \ddot{u}(n,t) = & \phi'(x(n+1,t)-x(n,t))\\&-\phi'(x(n,t)-x(n-1,t))- k [u(n,t) - v(n,t)]\,,\\
	m_2 \ddot{v}(n,t) = & k [u(n,t) - v(n,t)]\,,
	\label{1:gov0b}
\end{align}
where $u(n,t)$ and $v(n,t)$, respectively, denote the displacement of the $n^{\mathrm{th}}$ particle of mass $m_1$ and $m_2$ with respect to their equilibrium positions at time $t$. 

\begin{figure}[tb]
\centering
\subfloat[A solitary wave]{
\includegraphics{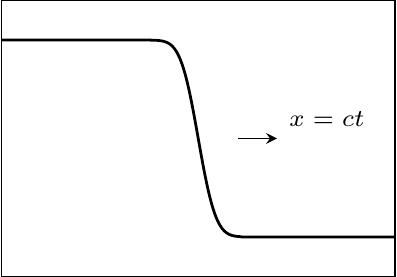}
}
\subfloat[A one-sided nanopteron]{
\includegraphics{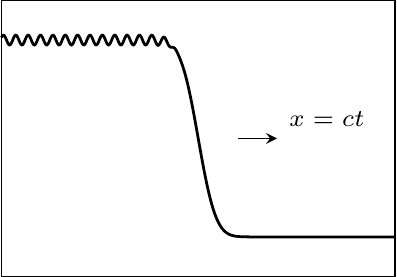}
}
\subfloat[A two-sided nanopteron]{
\includegraphics{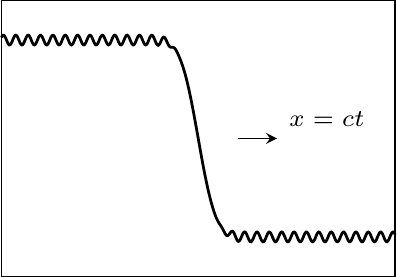}
}
\caption{Comparison of the profiles of (a) a standard solitary wave, (b) a one-sided nanopteron, and (c) a two-sided nanopteron that each propagates at speed $c$. The solitary wave is localized spatially, whereas the nanoptera have non-decaying oscillatory tails on (b) one side or (c) both sides of the wave front.
}
\label{f:nanoptera}
\end{figure}

Previous studies~\cite{Deng2021,Deng2022,Xu,Kim} have shown that traveling wave solutions in woodpile chains are often not localized solitary waves, but are instead a type of nonlinear wave known as a nanopteron. A nanopteron is the sum of a solitary wave a train of exponentially small oscillations that extend indefinitely in one or both directions of the central wave without decaying. These oscillations typically appear across Stokes curves, and must be studied using exponential asymptotic methods. In Figure \ref{f:nanoptera}, we illustrate examples of a solitary wave and one- and two-sided nanoptera.

\subsection{Rational approximation}

Rational approximation methods approximate a function as a ratio of a numerator and denominator polynomial. These methods are particularly accurate for functions which contain singularities in or near the approximation domain. Any approximation method that will be used for exponential asymptotics must be able to accurately describe singular behaviour; rational approximation methods are ideally suited for this purpose.

Historically, the most widely-used rational approximation method is Pad\'{e} approximation~\cite{pade}. The approximation is written in terms of the power series coefficients of the target function at a point. This method is well-suited to problems where the power series coefficients can be obtained exactly, and where we wish to approximate the solution accurately at a particular point. For many problems we wish to study, we wish to approximate the solution over a wider domain, and calculation of the power series coefficients requires the numerical approximation of higher-order derivatives of the solution. Due to these limitations, we apply the Pad\'{e} approximation only to Toda woodpile chains, for which the power series coefficients can be calculated exactly, for purposes of comparison.

More recently-developed rational approximation methods, such as the Antoulas-Anderson method~\cite{Antoulas}, AAA method~\cite{Nakatsukasa}, vector fitting~\cite{gustavsen1999rational}, rational Krylov fitting (RKFIT) method~\cite{Berljafa2015,Berljafa2017}, and iterative rational Krylov algorithm (IRKA)~\cite{Beattie2012,Gugercin2008}, avoid some of the challenges faced when applying the Pad\'{e} approximation. These methods typically fit a rational approximation to a set of data points distributed over a larger domain, with the pole locations determined by some least-squares optimization process. The discussion in~\cite{Nakatsukasa} contains a discussion regarding the accuracy and speed of these methods. For the purposes of this study, we will apply the AAA algorithm. 

The Antoulas-Anderson method involves interpolating every pair of points to determine a parametrization of all minimal-degree rational functions. The AAA method~\cite{Nakatsukasa} adapts this method to solve a least-squares minimization problem to approximate the solution on a set of support points across the domain. The AAA method has been applied to solve nonlinear eigenvalue problems~\cite{Lietaert}, represent conformal maps~\cite{Trefethen2020,Gopal1}, calculate rational minimax approximations~\cite{Filip}, and compute fractional diffusion~\cite{Hofreither}. In this paper, we will apply a method based on AAA approximation to obtain leading-order approximations to both Toda and woodpile nonlinear chains. The Pad\'{e} and AAA approximation methods are described in Section \ref{s:rational}.

\section{Exponential asymptotic method}
\label{s:exponential}

The methodology used in this study is very similar to that of~\cite{Lustri,Lustri1,Deng2021,Deng2022}. The explanation of the methodology is similar to the explanation contained in these previous studies.

We study the oscillatory tails of nanoptera in woodpile chains. We denote the mass ratio between light particles and heavy particles, with mass $m_2$ and $m_1$ respectively, by $\eta^2=m_2/m_1$. This system is singularly perturbed in the limit $\eta\to0$. The oscillatory tails are exponentially small in $\eta$.

Exponential asymptotic methods were developed in~\cite{Berry1988,Berry}, and applied to study Stokes' phenomenon in special functions. The key observation was that if an asymptotic series is truncated to minimise the error, the remainder term is exponentially small. By rescaling the problem to study the remainder term, it is possible to directly compute the exponentially small behaviour. These techniques were further developed in~\cite{BerryHowls1990} to further reduce the truncation error. See~\cite{Berry1991} for a  summary of these results.

The method applied in this study was developed in~\cite{Chapman,Daalhuis}. We write the solution $u$ to the singularly-perturbed governing equation as the power series
\begin{equation}\label{e:series_intro}
	u \sim \sum_{j=0}^\infty \eta^{r j}u_j \quad \mathrm{as} \quad \eta \rightarrow 0\,,
\end{equation}
where $r$ is the number of times that $u_{j-1}$ must be differentiated to obtain $u_j$. The leading-order solution, $u_0$, is obtained by setting the small parameter $\eta$ to be equal to zero, and solving the resultant equation.

We obtain an expression for $u_j$ by substituting the series~\eqref{e:series_intro} into the governing equation and matching terms at powers of $\eta$. To determine $u_j$ we need to differentiate earlier terms in the power series. If the leading-order solution $u_0$ contains singular points, the repeated differentiation guarantees that later terms in the series exhibit asymptotic behaviour known as ``factorial-over-power divergence"~\cite{Dingle}. 

To capture the factorial-over-power divergence, ~\cite{Chapman} propose an asymptotic ansatz for the terms $u_j$ in the limit $j\rightarrow\infty$, known as ``late-order terms",
\begin{equation}
	u_j\sim\frac{U\Gamma(r j+\gamma)}{\chi^{r j+\gamma}}\quad \mathrm{as} \quad j\to\infty\,.
\label{e:lateorder_intro}
\end{equation}
The parameter $\gamma$ is constant and $G$ and $\chi$ are functions of any independent variables but are independent of $j$. The function $\chi$ is known as the ``singulant".  It is equal to $0$ at each singularity of the leading-order solution $u_0$, ensuring that each $u_j$ is singular at the same locations as the leading-order solution. 

By substituting~\eqref{e:series_intro} and~\eqref{e:lateorder_intro} into the governing equation and matching orders of $\eta$, we can find $\chi$ and $G$. Through a local analysis of the solution near singular points, we can obtain  $\gamma$. Stokes curves originate at singular points of $u_0$, which satisfy $\chi = 0$, and follow curves on which $\chi$ is real and positive.

The series~\eqref{e:series_intro} is then truncated, with the heuristic from~\cite{Boyd1999} used to determine the optimal truncation point $N_{\mathrm{opt}}$. This gives
\begin{equation}
	u = \sum_{j=0}^{N_{\mathrm{opt}}-1} \eta^{r j}u_j+u_{\exp}\,,
\label{e:series_intro_1}
\end{equation}
where $u_{\mathrm{exp}}$ is the exponentially small truncation remainder. By substituting~\eqref{e:series_intro_1} into the governing equation, we can obtain a new equation for the exponentially small remainder term. Following~\cite{Daalhuis}, we can apply matched asymptotic expansions in the neighbourhood of the Stokes curve to find that the remainder has the form 
\begin{equation}
	g_{\exp}\sim\mathcal{S}G\e^{-\chi/\eta} \quad \mathrm{as} \quad \eta \rightarrow 0\,,
\label{e:stokes_intro}
\end{equation}
where $\mathcal{S}$ is known as the Stokes multiplier; it is a function of the independent variables in the problem. Away from the Stokes curve, the Stokes multiplier takes constant value, and~\eqref{e:stokes_intro} reduces to the standard WKB ansatz. In the neighborhood of the Stokes curve of width  $\mathcal{O}(\sqrt{\eta})$ with $\eta\to0$, the Stokes multiplier $\mathcal{S}$ undergoes a rapid change, which is known as Stokes switching. 

The leading-order behaviour $u_0$ plays several important roles in this process. Firstly, the terms $u_j$ are singular at singular points of $u_0$ in the complex plane with the singularity strength increasing with $j$, leading to factorial-over-power divergence. Secondly, Stokes curves in the solution originate at singular points of $u_0$. Finally, from \eqref{e:stokes_intro} we see that the exponential scaling of $u_{\mathrm{exp}}$ is determined by the appearance of $\chi$ in the exponent, and $\chi$ satisfies the condition that $\chi = 0$ at singular points of $u_0$.

The reliance of exponential asymptotic methods on the explicit calculation of $u_0$ causes challenges when studying systems for which $u_0$ cannot be calculated exactly. The purpose of this study is to determine whether numerical approximations to the leading-order behaviour -- in particular, the numerical analytic continuation of a simulated solitary wave solution --  can be used to produce an approximation, denoted $\hat{u}_0$, that is able to replicate exponential asymptotic results. 


\section{Rational approximation methods}
\label{s:rational}

In this section, we outline two methods for analytic continuation using rational function approximation that we will use as part of a numerical algorithm for finding the leading-order approximation $\hat{u}_0$. 

\subsection{Pad\'{e} approximation}\label{e:intro_Pade}
The $[N,M]$ Pad\'{e} approximant, denoted $[N,M](x)$ to a function with power series given by
\begin{align}
F(x)=\sum_{j=0}^\infty f_j x^j
\label{e:intro_coefficient}
\end{align}
is written explicitly in terms of the power series coefficients $f_j$ in~\eqref{e:intro_coefficient} as $[N, M](x) = P(x)/Q(x)$, where $P(x)$ and $Q(x)$ are the determinants of $N \times N$ matrix expressions given in \cite{pade}.
%
The $[N,N](\xi)$ rational approximation has the same power series as $F(x)$ up to order $N$. Increasing the dimension of the matrix $N$, and therefore the accuracy of the approximation, introduces more poles into the rational approximation.

For woodpile chains with Toda particle interactions, we can explicitly determine the leading-order solution exactly, and it is therefore straightforward to obtain $f_j$ exactly. For woodpile chains with power-law interactions, we cannot find an exact solution. Instead, we have a numerical leading-order solution. To obtain $f_j$, we need to calculate higher-order terms in the power series numerically. This is impractical, as taking differentiation magnifies accumulated numerical inaccuracy.

\subsection{The AAA approximation}

The Pad\'{e} approximation determines the numerator and denominator polynomials by evaluating the derivative of the function at a single point. We also consider exponential asymptotics using the AAA method, which is based on fitting the numerator and denominator polynomials at points sampled over a domain. We summarise the first presentation of this method, from \cite{Nakatsukasa}.

The AAA method approximates the function $f(z)$ as a rational expression,
\begin{align}
f(z) \approx\frac{n(z)}{d(z)}={\sum_{j=1}^m\frac{w_jf_j}{z-z_j}} \, \Bigg/ \, {\sum_{j=1}^m\frac{w_j}{z-z_j}}.
\label{e:intro_aaa}
\end{align}
The weights $w_j$ and the pole locations $z_j$ are found using an iterative method. The $m^{\mathrm{th}}$ iteration takes as input the values of $z_j$ for $j = 1, \ldots,m$. The first part of the iteration process produces the values of $w_j$ for $j = 1,\ldots, m$, and the second part produces $z_{m+1}$. 

The first part of the process involves solving a linear least-squares problem over a set of points in the restricted domain $Z^{(m)}=Z \setminus \{z_1,\ldots,z_{m}\}$; these points are labelled $Z_i^{(m)}$. The use of the restricted domain ensures that the rational approximation does not contain poles on the set. The function $f$ is interpolated, giving $f_1=f(z_1),\dots,f_m=f(z_m)$ at $z_1,\dots,z_m$. The weight vector $w =(w_1,w_2,\dots,w_m)^T$ is chosen to solve the least-squares problem of minimizing $\norm{fd-n}_{Z^{(m)}}$ subject to $\norm{w}_m=1$, where $\norm{\cdot}_{Z^{(m)}}$ is the discrete 2-norm over $Z^{(m)}$ and $\norm{\cdot}_m$. is the discrete 2-norm on $m$-vectors.
This is equivalent to solving the matrix problem
\begin{align}
\mathrm{minimize}\,\,\,\,
\norm{A^{(m)}w}_{M-m},\quad \norm{w}_m=1,
\label{e:AAA_intro}
\end{align}
where $A$ is a $(M-m)\times m$ matrix given by
\begin{align}
A^{(m)} = \begin{bmatrix}
\frac{F_1^{(m)}-f_1}{Z_1^{(m)}-z_1} & \dots         & \frac{F_1^{(m)}-f_m}{Z_1^{(m)}-z_m}         \\
\vdots   &  \ddots    & \vdots \\
\frac{F_{M-m}^{(m)}-f_1}{Z_{M-m}^{(m)}-z_1}         & \dots    & \frac{F_{M-m}^{(m)}-f_m}{Z_{M-m}^{(m)}-z_m}            \\
\end{bmatrix}.
\end{align}
This minimization problem can be solved efficiently using the singular value decomposition (SVD), and taking $w$ as the final right singular vector in a reduced SVD. 

Finally, the value of $z_{m+1}$ is determined for the next iteration. It is found by choosing the value of $z_{m+1}$ that maximizes the quantity
\begin{align}
\sum_{j=1}^m \frac{w_j f(z_{m+1})}{z_{m+1}-z_j}-
\sum_{j=1}^m \frac{w_j f_j}{z_{m+1}-z_j}.
\end{align}
This process is repeated until the $L^{\infty}$ norm of the difference between the input function and the rational approximation is smaller than a specified tolerance.

One potential obstacle in applying rational approximation methods is that these methods produce meromorphic functions; all singularities in the approximation are isolated poles. If the leading-order behaviour contains branch points, the solution will not be accurately approximated in some local vicinity of the pole. In previous studies of AAA approximation~\cite{Gopal,Trefethen2020,Trefethen2021}, it was noted that the presence of branch points in the solution creates an exponential clustering of poles (and zeroes) in the approximation around the branch points of the target function, which accurately approximate the effects of a branch cut. A goal of this study is to determine whether this pole accumulation is capable of accurately predicting exponentially small behaviour in Toda woodpile chains, whose leading-order solution contains logarithmic branch cuts.

\section{Woodpile chains with Toda interactions}
\label{s:woodToda}

The scaled governing equations for a woodpile chain governed by a Toda interaction potential \eqref{e:potentialToda}, with $a = 1$ and $b=1$, are given by
\begin{align}\label{1:gov1Toda}
	\ddot{u}(n,t) &= \e^{-[u(n,t)-u(n-1,t)]} - \e^{-[u(n+1,t)-u(n,t)]}-k [{u}(n,t) - {v}(n,t)]\,,\\
\eta^2 \ddot{{v}}(n,t) &= {k} [{u}(n,t)-{v}(n,t)]\,.
\label{1:gov2Toda}
\end{align}
In the limit $\eta \rightarrow 0$ we expand $u(x,t)$ and $v(x,t)$ as a power series in $\eta^2$, giving
\begin{align}
	u(n,t)\sim\sum_{j=0}^\infty \eta^{2j}u_j(n,t)\,, \quad v(n,t)\sim\sum_{j=0}^\infty \eta^{2j}v_j(n,t)\,.
\label{1:asympseriesToda}
\end{align}
Substituting the series expression~\eqref{1:asympseriesToda} into~\eqref{1:gov2Toda} and matching at the leading order in the limit $\eta\to0$, we find that $u_0(n,t) = v_0(n,t)$, and
\begin{align}
\ddot{u}_0(n,t) &= \e^{-[u_0(n,t)-u_0(n-1,t)]} - \e^{-[u_0(n+1,t)-u_0(n,t)]}\,.
\label{1:woodpile12Toda}
\end{align}
We introduce a co-moving frame $\xi=n-ct$, where $c$ is the solitary wave velocity. In the co-moving frame,~\eqref{1:woodpile12Toda} can be written as
\begin{align}
c^2{u}''_0(\xi) &= \e^{-[u_0(\xi)-u_0(\xi-1)]} - \e^{-[u_0(\xi+1)-u_0(\xi)]}\,,
\label{1:woodpile12Toda_xi}
\end{align}
where prime denotes derivative with respect to $\xi$. 

Equation~\eqref{1:woodpile12Toda_xi} has an exact solitary-wave solution, which we present in \eqref{e:LO_precise}. This exact solution will be applied to determine the form of the exponentially small oscillations in the nanopteron solutions for the woodpile system using exponential asymptotics. We will then use different numerical approaches to approximate $u_0$, giving an approximate leading-order solution (denoted $\hat{u}_0$), and compare the output of exponential asymptotic analyses based on these approximate solutions.

The leading-order solution is singular at points in the complex plane, and these singularities occur in complex-conjugate pairs. We denote the singularity locations as $\xi_s$ and $\xi_s^*$, where $\mathrm{Im}(\xi_s)>0$. Each singularity pair is connected by a Stokes curve, which switches on an exponentially small contribution. 

Assume that $u_0$ is singular at $\xi = \xi_s$, and that the singularity behaviour is locally either logarithmic, so that $u_0 \sim \mu \log(\xi - \xi_s)$ as $\xi \to \xi_s$ for some $\mu$, or a pole or branch point, so that $u_0 \sim \mu(\xi - \xi_s)^{\nu}$ as $\xi \to \xi_s$ for some $\mu$ and $\nu$. This local behaviour is sufficient to determine the exponentially small contribution that appears across the Stokes curve; this analysis is presented in Appendix \ref{A:ExpAsymp}. The exponentially small solution contribution to $v(\xi)$, denoted $v_{\mathrm{exp}}$, is given by
\begin{equation}
	v_{\mathrm{exp}} \sim\frac{2|V|   \pi}{\eta^{\nu}}
\exp\left(-\frac{\sqrt{k}\mathrm{Im}(\xi_s)}{c\eta}\right)
\cos\left(\frac{\sqrt{k}(\xi-\mathrm{Re}(\xi_s))}{c\eta}-\phi\right)\quad \mathrm{as} \quad \eta \rightarrow 0\,,
\label{e:woodpile_sn}
\end{equation}
where $V = \mu$ if the singularity is logarithmic, and $V = \mathrm{i} \mu \sqrt{k}/c$ if the solution is a pole or branch point. In the case of a logarithmic singularity, we take $\nu = 0$. The quantity $\phi$ is given by $\pi/2$ if the singularity is logarithmic, and the phase of $V$ if the singularity is a pole or branch cut. The term $u_{\mathrm{exp}}$ can be obtained directly from $v_{\mathrm{exp}}$, but we omit these details here.

This contribution is present for $\mathrm{Re}(\xi) > \mathrm{Re}(\xi_s)$, which corresponds to the region behind the leading-order solitary wave. The amplitude of the oscillations decays exponentially as $\mathrm{Im}(\xi_s)$ increases. The dominant oscillations are therefore those associated with the singularity pairs that are nearest to the real axis, which are exponentially larger in the small $\eta$ limit than singularity pairs which are further away.

\subsection{Exact leading-order solution}\label{S:exactToda}

\begin{figure}[tb]
\centering
\subfloat[Leading-order solution $u_0(\xi)$ for $\kappa = 1$]{
\includegraphics[scale=1.00]{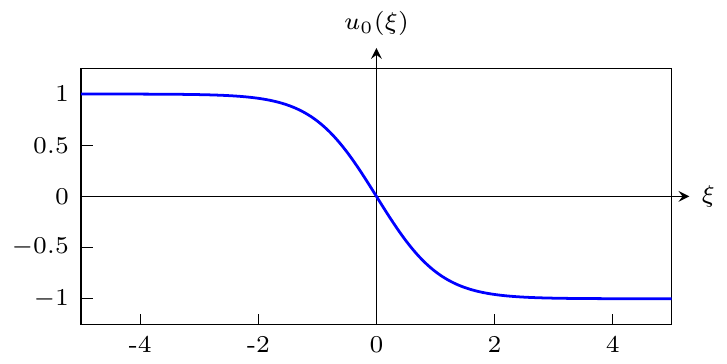}
}
\subfloat[$\mathrm{Re}(u_0)$]{
\includegraphics[scale=1.00]{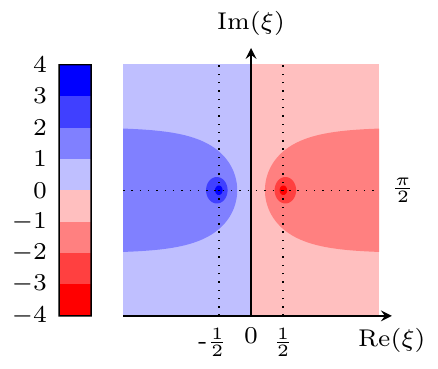}
}
\subfloat[$\mathrm{Im}(u_0)$]{
\includegraphics[scale=1.00]{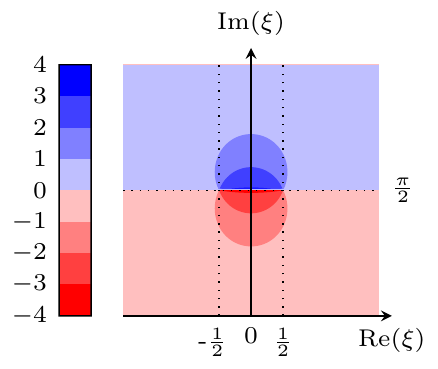}
}

\subfloat[Stokes Curves]{
\includegraphics[scale=1.00]{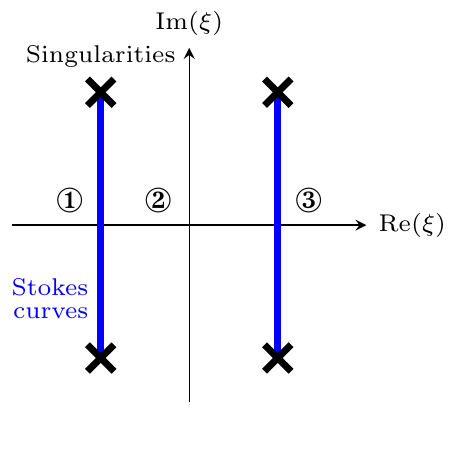}
}
\subfloat[Comparison of Results]{
\includegraphics[scale=1.00]{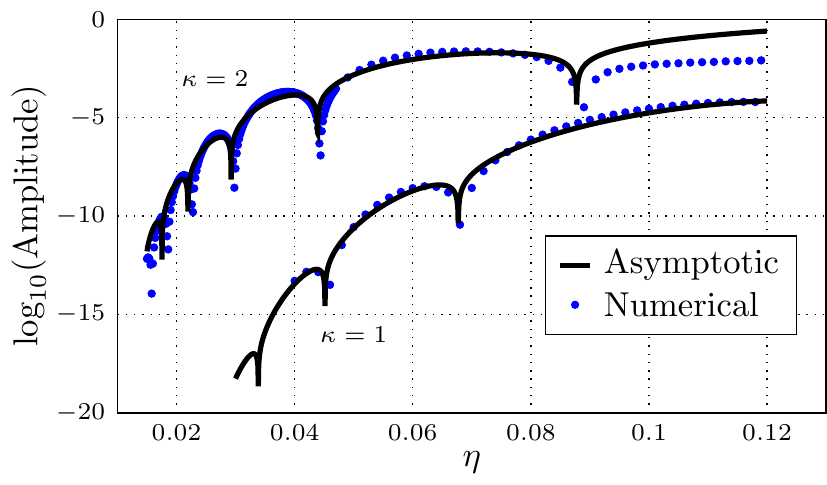}
}
\caption{(a) The exact leading-order wave $u_0$ as a function $\xi$ for $\kappa=1$. (b)--(c) The behaviour of $u_0$ for $\xi \in \mathbb{C}$ in the upper half-plane. (d) The Stokes curves that connect the singularity pairs. As Stokes curves are crossed from right to left, exponentially small oscillations appear in the solution. In \ding{172} there are no oscillations. In \ding{173}, there is one oscillatory contribution. In \ding{174}, both oscillatory contributions are present. When these oscillations have opposite phase, they cancel. Region \ding{174} contains no oscillations in this case, and the solution is a solitary wave. (e) Comparison between asymptotic predictions and simulations of the oscillation amplitudes for $\kappa=1$ and $\kappa=2$ respectively, with $k=1$. The asymptotic predictions accurately predict the oscillation amplitude and the values of $\eta$ at which the oscillations disappear.}
\label{f:SW_Toda}
\end{figure}

Equation~\eqref{1:woodpile12Toda} is a scaled version of the integrable Toda chain and may be solved exactly, as in~\cite{Toda}. Recalling that $u_0 = v_0$, the leading-order solution to the system~\eqref{1:gov1Toda}--\eqref{1:gov2Toda} is given by
\begin{align}
u_0(n,t)=\log\left(\frac{\cosh(\kappa(n-1)-\beta t)}{\cosh(\kappa n-\beta t)}\right)\,,\quad u_0(n,t) = v_0(n,t)\,,
\label{e:LO_precise}
\end{align}
where $\kappa$ is the soliton parameter, $\beta=\sinh\kappa$  and the velocity of the soliton is given by $c=\sinh\kappa/\kappa$. We write this in terms of an offset co-moving frame, given by $\xi = n - ct - 1/2$, where the offset is introduced to make the wave symmetric about $\xi = 0$. The solution in this frame is given by
\begin{align}
u_0(\xi)=\log\left(\frac{\cosh(\kappa( \xi-1/2))}{\cosh(\kappa (\xi+1/2))}\right)\,,\quad u_0(\xi) = v_0(\xi)\,.
\label{e:LO_precise_comoving}
\end{align}
The leading-order solution $u_0$ is shown in Figure~\ref{f:SW_Toda}(a).

If we analytically continue the leading-order solutions $u_0$ and $v_0$ so that $\xi \in \mathbb{C}$, we see that these functions are singular at a set of points described by
\begin{align}
\xi_{1,\pm,j} = -\tfrac{1}{2}\pm\tfrac{\mathrm{i}\pi}{2\kappa}(2j+1)
\quad\mathrm{and}\quad
\xi_{2,\pm,j} = \tfrac{1}{2}\pm\tfrac{\mathrm{i}\pi}{2\kappa}(2j+1),
\label{e:singularity_precise}
\end{align}
where $j$ is a non-negative integer. The behaviour of $u_0$ for $\xi \in \mathbb{C}$ in the upper half-plane is shown in Figure~\ref{f:SW_Toda}(b)--(c). These contour plots show the real and imaginary part of $u_0(\xi)$, and depict the two singularities nearest to the real axis. Recall that the exponentially small behaviour is dominated exponentially as $\eta \to 0$ by contributions from the singularities that are nearest to the real axis.

These singularities are connected by Stokes curves, shown in Figure~\ref{f:SW_Toda}(d). There are two exponentially small contributions; one appears on the left-hand side of the curve $\mathrm{Re}(\xi) = -1/2$, and one appears on the left-hand side of the curve $\mathrm{Re}(\xi) = 1/2$. The oscillatory behaviour in the nanopteron solution is therefore a superposition of the two sets of exponentially small oscillations.

From~\eqref{e:woodpile_sn}, we determine that the exponentially small oscillations generated by the singularity pair located at $\xi = \xi_{2,\pm,j}$ have the form
\begin{equation}
	v_{\mathrm{exp}} \sim 2\pi
\exp\left(-\frac{\sqrt{k}\pi(2j+1)}{2\kappa c\eta}\right)
\sin\left(\frac{\sqrt{k}(\xi-1/2)}{c\eta}\right)\quad \mathrm{as} \quad \eta \rightarrow 0\,.
\label{e:woodpile_sn1}
\end{equation}
The exponentially small oscillations generated by singularity pair located at $\xi = \xi_{1,\pm,j}$ have the form from \eqref{e:woodpile_sn1}, with an offset in $\xi$ of $+1/2$ rather than $-1/2$. From these two expressions, it is clear that the dominant oscillations are those obtained by choosing $j = 0$. Taking the sum of these oscillations and simplifying them allows us to determine the amplitude of the combined oscillatory behaviour, given by
\begin{equation}
\mathrm{Amplitude} \sim4\pi\exp\left(-\frac{\sqrt{k}\pi(2j+1)}{2\kappa c\eta}\right)\sin\left(\frac{\sqrt{k}}{2c\eta}\right).
\label{e:toda_amp}
\end{equation}

A comparison between the asymptotic prediction of the amplitude and numerical results calculated using the numerical scheme from \cite{Deng2021, Deng2022} is shown in Figure~\ref{f:SW_Toda}(e). The asymptotic predictions show strong agreement with the numerical simulations, especially for small values of $\eta$. The asymptotic analysis is able to accurately predict values of the mass ratio at which the oscillations disappear entirely (which appear in Figure~\ref{f:SW_Toda}(e) as sharp negative cusps); these correspond to solitary-wave solutions.


\subsection{Tanh-fitted leading-order solution}
\label{ss:Toda_tanh}

\begin{figure}[tb]
\centering
\subfloat[Leading-order approximation $\hat{u}_0(\xi)$ for $\kappa = 1$]{
\includegraphics[scale=1.00]{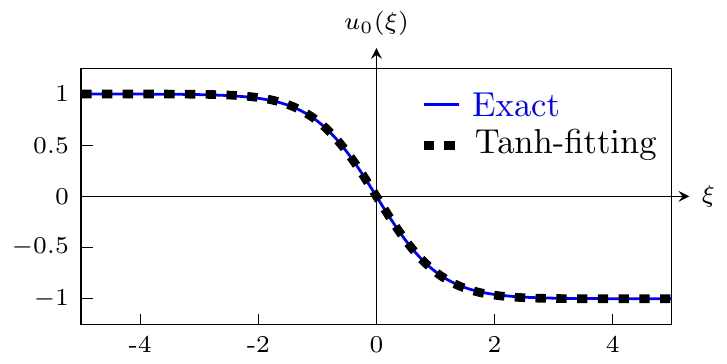}
}
\subfloat[$\mathrm{Re}(\hat{u}_0)$]{
\includegraphics[scale=1.00]{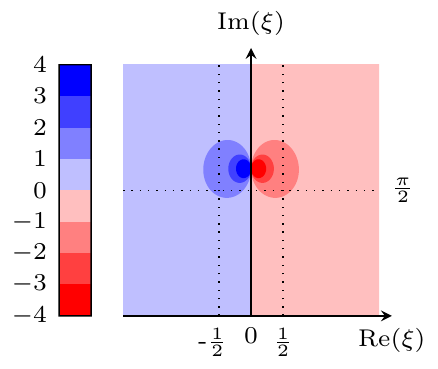}
}
\subfloat[$\mathrm{Im}(\hat{u}_0)$]{
\includegraphics[scale=1.00]{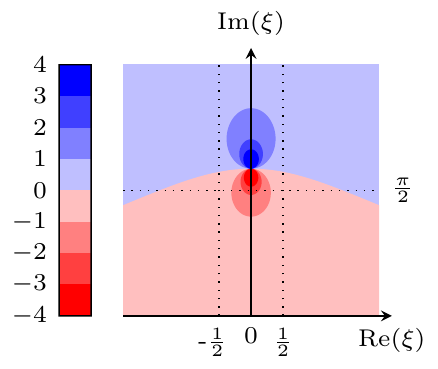}
}

\subfloat[Leading-order approximation $\hat{u}_0(\xi)$ for $\kappa = 2$]{
\includegraphics[scale=1.00]{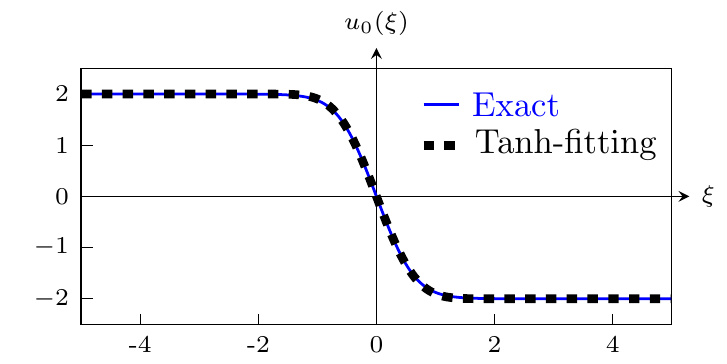}
}
\subfloat[$\mathrm{Re}(\hat{u}_0)$]{
\includegraphics[scale=1.00]{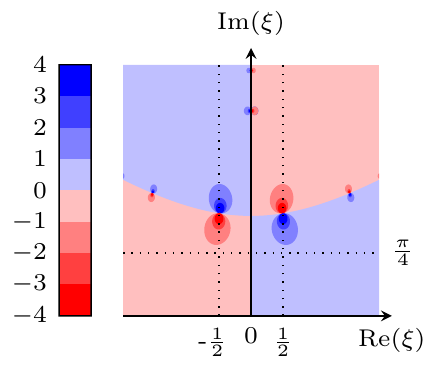}
}
\subfloat[$\mathrm{Im}(\hat{u}_0)$]{
\includegraphics[scale=1.00]{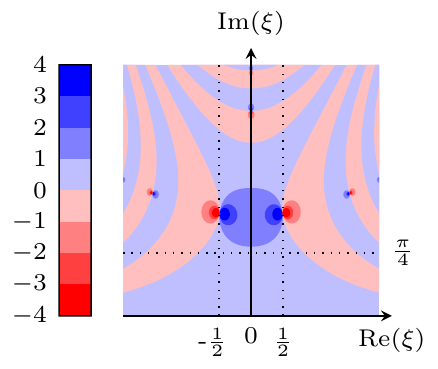}
}
\caption{A comparison between the tanh-fitted leading order solution $\hat{u}_0$ and the exact leading order solution $u_0$ for (a) $\kappa = 1$, and (d) $\kappa = 2$. The analytic continuation of $\hat{u}_0$ is shown in (b)--(c) for $\kappa = 1$ and (e)--(f) for $\kappa = 2$. For $\kappa = 1$, $\hat{u}_0$ contains one dominant singularity pair. For $\kappa = 2$, $\hat{u}_0$ contains two equally dominant singularity pairs. Hence, $\hat{u}_0$ cannot produce solutions for which the oscillations cancel if $\kappa = 1$, but this is possible for $\kappa = 2$.}
\label{f:amplitude_tanh}
\end{figure}

In~\cite{sen2001}, the authors note that solitary waves often have a kink shape, and proposed a method of approximating solitary waves by fitting a tanh function to the solution. The authors assume solitary waves in Hertz chains can be approximated as $u_0 \approx \hat{u}_0$, where $\hat{u}_0$ has the form
\begin{align}
\hat{u}_0(\xi) = -\frac{A}{2}\tanh(f(\xi)),\quad f(\xi)=\sum_{n=0}^N C_{2n+1}\xi^{2n+1}, \quad \xi=n-c_A t,
\label{e:senmanciu}
\end{align}
where $A$ is the amplitude of the solitary wave, $c_A$ is the velocity of the solitary wave with amplitude $A$, $N$ is a positive integer, $\xi$ defines the co-moving frame and the coefficients $C_{2n+1}$ are determined by numerically approximating higher-order derivatives of the leading-order behaviour and equating the results with an analytic expression for the corresponding derivatives of \eqref{e:senmanciu}. This approximation is appropriate a solitary wave profile for which $u_0(\xi) \to \mp A/2$ as $\xi \to \pm \infty$. It is clear from Figure~\ref{f:SW_Toda}(a) that the Toda solitary wave falls into this category.

For $\kappa=1$, we obtain $C_1=0.92423$, $C_3=0.020876$ and a negative value for $C_5$. For $\kappa=2$, we obtain $C_1=1.5232$, $C_3=0.32505$, and $C_5<0$. If negative coefficients are included in the approximation, the profile behaves correctly in the vicinity of $\xi = 0$, but the wave profile is not monotonic and can tend to incorrect values as $|\xi| \to \infty$. There is a limit to the approximation accuracy that can be obtained using this method, and it cannot be continued indefinitely. Hence, we only consider approximations using coefficients up to $C_3$. 

We show a comparison the approximation $\hat{u}_0$ and the true leading-order solution $u_0$ in Figure~\ref{f:amplitude_tanh}(a) and (d) for $\kappa=1$ and $\kappa=2$. The behaviour of $\hat{u}_0$ in the complex plane is presented in Figure~\ref{f:amplitude_tanh}(b)--(c) for $\kappa=1$ and (e)--(f) for $\kappa=2$. These figures show a difference between the two cases. For $\kappa = 1$ there is a single dominant singularity pair, located at $\xi = \pm 1.84035\mathrm{i}$. For $\kappa = 2$, there are two equally dominant singularity pairs, located at $\xi = -0.4923\pm 1.2817\mathrm{i}$ and $\xi = 0.4923\pm 1.2817\mathrm{i}$. 

In fact, there exists a critical value $\kappa_c\approx1.56$ for $\kappa$. If $\kappa<\kappa_c$, then $\hat{u}_0$ contains one singularity pair closest to the real axis; when $\kappa>\kappa_c$, the single singularity pair splits into two distinct pairs. Oscillations can only cancel if there are two contributions of the same amplitude. Hence, the oscillations can only vanish if $\kappa$ exceeds this critical value. From the results in Figure \ref{e:toda_amp}, we know this is not a true feature of the asymptotic approximation, but rather an inaccuracy introduced by the approximation method.

\begin{figure}
    \centering
    \includegraphics[scale=1.00]{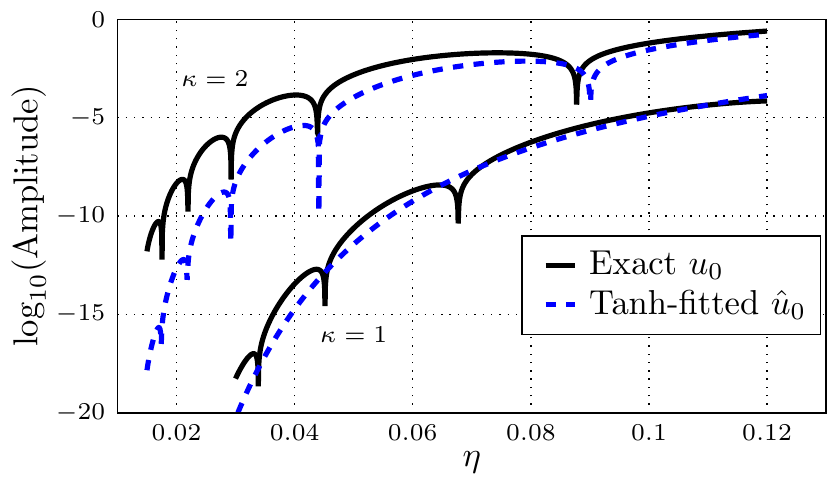}
    \caption{Comparison of asymptotic predictions based on the approximate leading-order $\hat{u}_0$ and predictions based on $u_0$. For $\kappa =1$, there are no values of $\eta$ which result the oscillations vanishing, while for $\kappa =2$, the predicted values of $\eta$ at which the oscillations vanish are accurate. In both cases, the amplitude is not accurately predicted by the exponential asymptotic method.}
       \label{f:amplitude_tanh1}
\end{figure}

Using~\eqref{e:woodpile_sn} and the local behaviour of $\hat{u}_0$ near singularities, the amplitude of the oscillations for $\kappa = 1$ is given by
\begin{align}
\label{e:amp_kappa1}
\mathrm{Amplitude} \sim \frac{1.4043 \pi\sqrt{k}}{\eta c_A}
\exp\left(-\frac{1.84035\sqrt{k}}{c_A\eta}\right).
\end{align}
For $\kappa=2$,  the amplitude of the oscillations is given by
\begin{align}
\label{e:amp_kappa2}
\mathrm{Amplitude} \sim \frac{1.6120  \pi\sqrt{k}}{\eta c_A}
\exp\left(-\frac{1.2817\sqrt{k}}{c_A\eta}\right)
\cos\left(\frac{0.4923\sqrt{k}}{c_A\eta}-1.4435\right).
\end{align}
We show a comparison between the amplitude based on the approximation $\hat{u}_0$ and that based on the exact leading-order solution $u_0$ in Figure~\ref{f:amplitude_tanh1}. For $\kappa=1$, which is below the critical value, the asymptotic prediction cannot identify values of $\eta$ at the oscillations vanish, and the approximation underestimates the amplitude.  For $\kappa=2$, the asymptotic prediction identifies the values of $\eta$ at which the oscillations vanish, but the amplitude deviates significantly from numerical simulations. 

\subsection{Pad\'{e}-approximated leading-order solution}
\label{ss:Toda_Pade}

\begin{figure}
\centering
\includegraphics[scale=1.00]{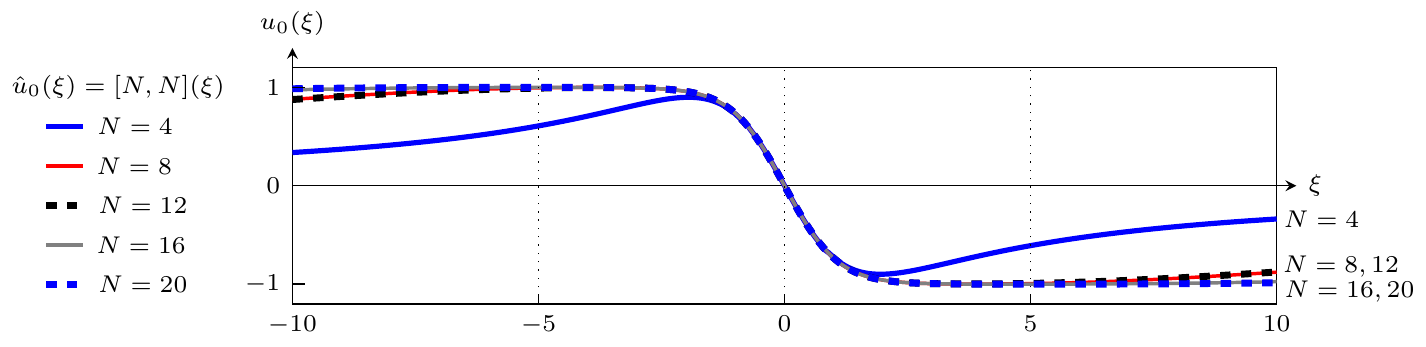}
\caption{Pad{\'e} approximants with different values of $N$, which is the order of the numerator and denominator polynomials. The size of the region in which the approximant agrees well with the exact solution increases as $N$ increases. For $N \geq 8$, the approximant is accurate in the region containing the solitary wave, although the approximants generated using $N = 8$, $16$ do differ visibly from the exact solution for $|\xi| > 5$. }
\label{f:pade_approximant}
 \end{figure}

In this section we use the Pad\'{e} approximant described in~\eqref{e:intro_Pade} to approximate the leading-order solution. As we are approximating a function which tends to constant values at $\pm\infty$, we expect that the degree of the polynomial in the numerator will be close to that in the denominator, so we take $N=M$. 

In Figure~\ref{f:pade_approximant} we show that the size of the region in which the Pad\'{e} approximant accurately predicts the solution behaviour increases with $N$. We wish to select an approximation that accurately describes the behaviour of the solution in the region where the wave varies. From Figure~\ref{f:pade_approximant}, we see that choosing $N = 8$ is accurately describes the solitary wave in $\xi \in [-5,5]$. We could choose higher values of $N$, but the subsequent analysis does not change significantly.

\begin{figure}[tb]
\centering
\subfloat[Exact leading-order solution $u_0(\xi)$ for $\kappa = 1$]{
\includegraphics[scale=1.00]{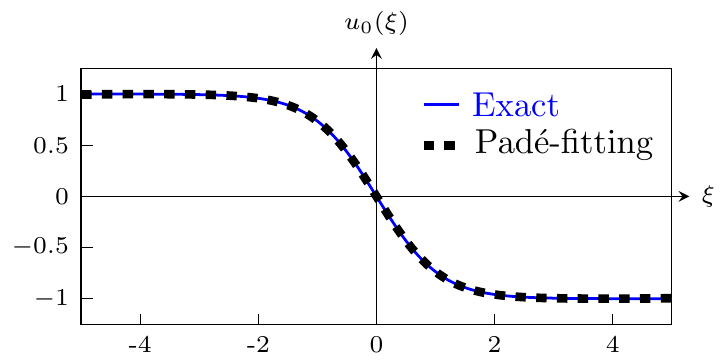}
}
\subfloat[$\mathrm{Re}(\hat{u}_0)$]{
\includegraphics[scale=1.00]{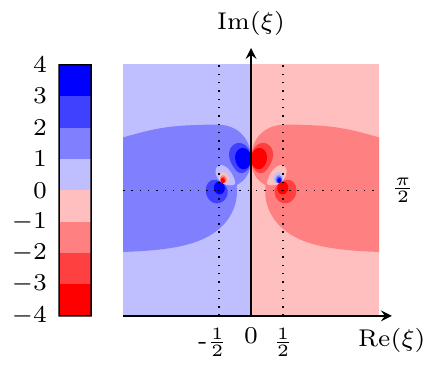}
}
\subfloat[$\mathrm{Im}(\hat{u}_0)$]{
\includegraphics[scale=1.00]{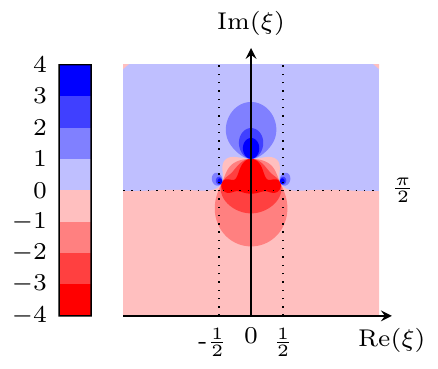}
}

\subfloat[Stokes Curves]{
\includegraphics[scale=1.00]{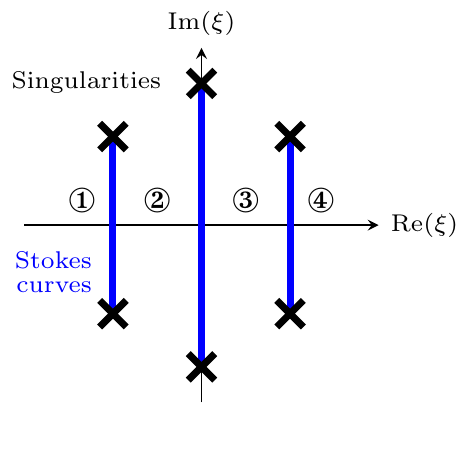}
}
\subfloat[Comparison of Results]{
\includegraphics[scale=1.00]{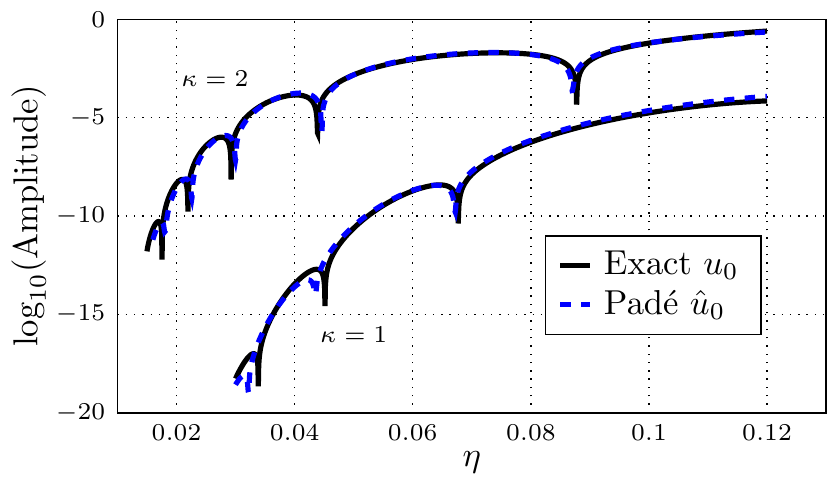}
}
    \caption{(a) Approximate leading-order wave $\hat{u}_0$ for $\kappa=1$. (b)--(c) The behaviour of $\hat{u}_0$ for $\xi \in \mathbb{C}$ in the upper half-plane. (d) The Stokes curves that connect the singularity pairs. In \ding{172} there are no oscillations. In \ding{173}, there is one oscillatory contribution. In \ding{174}, a second oscillatory contribution appears which has different amplitude to the first. In \ding{175}, a third oscillatory contribution is switched on. When the first and third oscillations are in opposite phase, they cancel. There are still oscillations present due to the second Stokes curve, but they are smaller than the contributions from the dominant poles. (e) Comparison between asymptotic predictions for $u_0$ and $\hat{u}_0$ for $\kappa=1$ and $\kappa=2$. The asymptotic predictions accurately predict both the amplitude and the values of $\eta$ at which the dominant oscillations vanish.}
    \label{f:amplitude_Pade}
\end{figure}

For $\kappa=1$, the Pad\'{e} approximant is used to generate $\hat{u}_0$, given by
\begin{align}
\hat{u}_0(\xi) = [8,8](\xi)= -\frac{0.9242\xi+0.5441\xi^3 +
 0.1132\xi^5+0.005663\xi^7}{1+ 0.8509\xi^2 + 0.2742\xi^4 +
 0.03295\xi^6+0.0004223\xi^8}.
\label{e:Pade_approx}
\end{align}
This approximation is shown in Figure~\ref{f:amplitude_Pade}(a), and the analytic continuation is shown in Figure~\ref{f:amplitude_Pade}(b)--(c). The solution contains two singularity pairs closest to the real axis, located at $\xi = -0.4559 \pm 1.6651\mathrm{i}$ and $\xi = 0.4559 \pm 1.6651\mathrm{i}$. The solution also contains subdominant pairs, one of which can be seen in Figure~\ref{f:amplitude_Pade}(b)--(c). These pairs will generate Stokes curves, shown in Figure~\ref{f:amplitude_Pade}(d). The oscillations will be dominated by contributions from the dominant pair, but the remaining poles will also switch on behaviour that does not necessarily cancel. The amplitude will nearly cancel at particular values of $\eta$, but the cancellation will not be perfect.

Using \eqref{e:woodpile_sn}, the amplitude of the oscillations generated by $\hat{u}_0$ is 
\begin{align}
\label{e:amp_Pade}
\mathrm{Amplitude} \sim \frac{1.311  \pi\sqrt{k}}{\eta c_A}
\exp\left(-\frac{1.6651\sqrt{k}}{c_A\eta}\right)
\cos\left(\frac{0.4559\sqrt{k}}{c_A\eta}+2.0766\right).
\end{align}
This expression does not contain the exponentially subdominant oscillations generated by the subdominant pole pairs. Hence, we don't expect the amplitude to truly disappear at values of $\eta$ for which this expression is equal to zero, as the subdominant contributions will not cancel at these points.

For $\kappa=2$, the Pad\'{e} approximant is used to generate $\hat{u}_0$, given by
\begin{align}
\hat{u}_0(\xi) = [8,8](\xi)= -\frac{ 3.04638 \xi + 259.567 \xi^3 + 182.411 \xi^5 + 160.653 \xi^7}{85.7651 \xi^2 + 107.787 \xi^4 + 102.879 \xi^6 + 25.8326  \xi^8}.
\label{e:Pade_approx_kappa2}
\end{align}
The singularities closest to the real axis are located at $\xi = -0.5205 \pm 0.08846\mathrm{i}$ and $\xi = 0.5205 \pm 0.8846\mathrm{i}$. The amplitude of the oscillations is given by \eqref{e:woodpile_sn} as
\begin{align}
\label{e:amp_Pade_kappa2}
\mathrm{Amplitude} \sim \frac{1.1690 \pi\sqrt{k}}{\eta c_A}
\exp\left(-\frac{0.8846\sqrt{k}}{c_A\eta}\right)
\cos\left(\frac{0.5205\sqrt{k}}{c_A\eta}+1.4161\right).
\end{align}
In Figure~\ref{f:amplitude_Pade}(e), the oscillation amplitude generated by $\hat{u}_0$ for $\kappa = 1$ and $\kappa = 2$ are compared with the oscillation amplitude generated by the exact solution $u_0$. These amplitudes were generated using all singularity pairs in the solution, rather than just the dominant pairs. In each case, the approximation accurately captures the asymptotic behaviour generated by the exact leading-order. It is able to correctly predict both the amplitude of the oscillations, and the values of $\eta$ at which the oscillations appear to cancel. This is not true cancellation, as there are still exponentially subdominant oscillations in the solution; however, these contributions are so small that they are not visibly apparent in the figure.

These results indicate that using rational approximation to perform numerical analytic continuation produces approximations that are useful for exponential asymptotics. Computing the Pad{\'e} approximation requires finding Taylor coefficients of the leading-order solution. For the woodpile chain with Toda interactions, these coefficients are straightforward to obtain. For a woodpile chain with power-law interactions, these coefficients can only be obtained numerically by approximating higher-order derivatives at the point $\xi = 0$ numerically. 

\subsection{AAA-approximated leading-order solution}
\label{ss:Toda_AAA}

Motivated by the accurate results produced by Pad\'{e} approximation, we formulate a method based on rational approximation for $u_0$ using the AAA algorithm; an implementation of this algorithm is included in the Chebfun \textsc{Matlab} package~\cite{Driscoll}. The first step is to compute the leading-order solitary wave $u_0$ numerically by exciting a monatomic Toda lattice with a velocity impulse. The second step is to use the AAA algorithm to obtain a rational approximation for the leading-order wave, $\hat{u}_0$.

\subsubsection{Numerical Leading-Order Solution}

For the numerical step, we use the velocity Verlet algorithm~\cite{Verlet,Allen} to predict the particle behaviour. This is a symplectic integrator that is designed to conserve the energy of a system. The velocity Verlet algorithm uses the discretization
\begin{equation}
	\vec{x}_{n+1}=\vec{x}_n+\vec{v}_n\Delta t+\frac{1}{2} \vec{a}_{n}(\Delta t)^2\,,  \quad  \vec{v}_{n+1}=\vec{v}_n+\frac{1}{2}(\vec{a}_n+\vec{a}_{n+1})\Delta t\,,
\end{equation}
where the time step is given by $\Delta t=t_n-t_{n-1}$, and $\vec{x}_n$, $\vec{v}_n$, and $\vec{a}_n$ are vector quantities that encode the displacements, velocities, and accelerations of the particles at time $t=t_n$. In a woodpile chain, the acceleration $\vec{a}_n$ can be obtained directly from $\vec{x}_n$. Each simulation contains a large number of particles on a periodic domain. The domain information and the size of the time steps is presented in Table~\ref{t:parameter}.
 \begin{table}[htbp]
 \caption{Parameters for simulations of leading-order solitary wave in the Toda woodpile chain.}
 \label{t:parameter}
 \centering
 \begin{tabular}{|c| c| c| c|} \hline
    $\kappa$ & Number of particles &  $\Delta t$ & Velocity impulse  \\ \hline
      1 & 300 &  0.01 & $\vec{v}_{61}(0) = 2.3525$ \\  \hline
      2 & 300 &  0.01 & $\vec{v}_{61}(0) = 7.2560$  \\ \hline
\end{tabular}
\end{table}

\begin{figure}[tb]
    \centering
    \includegraphics[scale=1.00]{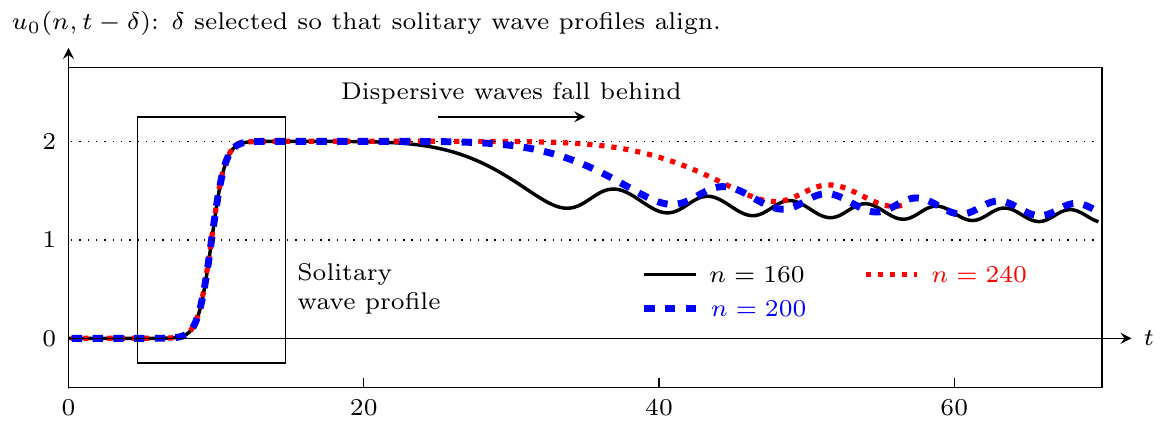}
    \caption{
    Displacement of particles $n =$ 160, 200, and 240 as a function of time. The time $t$ is offset by a shift ($\delta$) so that the solitary wave profiles are overlaid. The solitary wave profile in simulations is shifted up by half the amplitude, as the particles start with zero displacement. The solitary wave travels faster than the dispersive waves, and the two wave contributions seperate. Eventually, the seperation is large enough that we can isolate the solitary wave, without any measurable effects due to dispersive waves.}
    \label{f:particlemotion}
\end{figure}

Initially, all particles in the chain are at rest. At the initial time step, a velocity impulse is applied to the particle at $n = 61$. The velocity impulse generates a solitary wave and dispersive waves. The dispersive waves travel slower than the solitary wave, eventually seperating entirely. This is illustrated in Figure~\ref{f:particlemotion}, which shows the motion of the particles at $n = 160$, $200$, and $240$. As the distance from the initial impulse grows, the seperation between the solitary wave and the dispersive waves increases. At $n = 161$, the seperation is sufficient that we can isolate the numerical approximation to the solitary wave without capturing any effects due to the dispersive waves. We tune the magnitude of the velocity impulse to obtain solitary waves with amplitude $4$ and $2$, corresponding to $\kappa=2$ and $\kappa=1$. The parameters are given in Table~\ref{t:parameter}.

\begin{figure}[tb]
\centering
\subfloat[AAA-approximated $\hat{u}_0(\xi)$ for $\kappa = 1$]{
\includegraphics[scale=1.00]{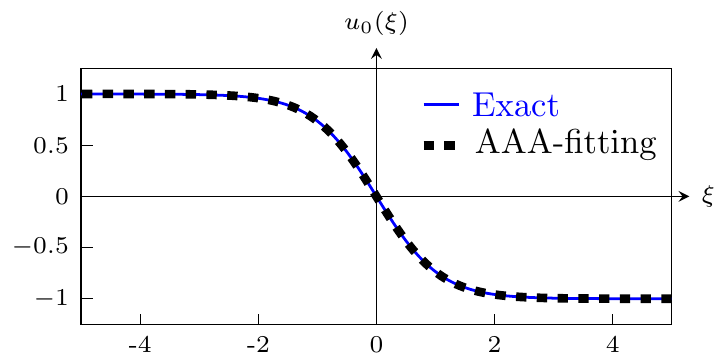}
}
\subfloat[$\mathrm{Re}(\hat{u}_0)$]{
\includegraphics[scale=1.00]{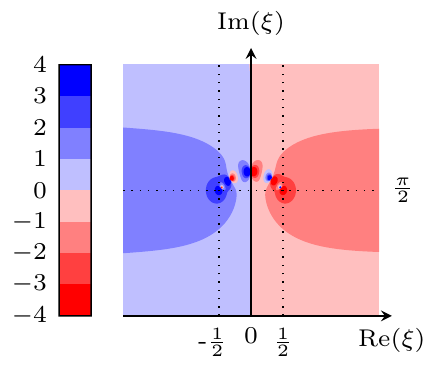}
}
\subfloat[$\mathrm{Im}(\hat{u}_0)$]{
\includegraphics[scale=1.00]{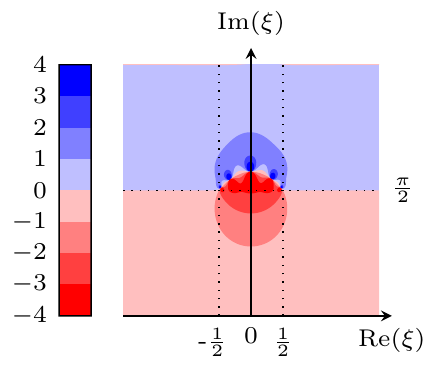}
}

\subfloat[AAA-approximated $\hat{u}_0(\xi)$ for $\kappa = 2$]{
\includegraphics[scale=1.00]{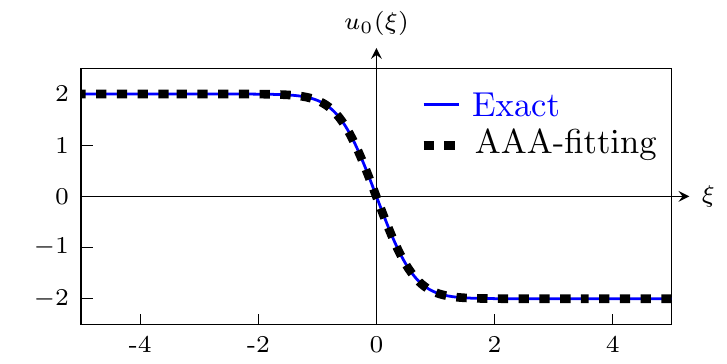}
}
\subfloat[$\mathrm{Re}(\hat{u}_0)$]{
\includegraphics[scale=1.00]{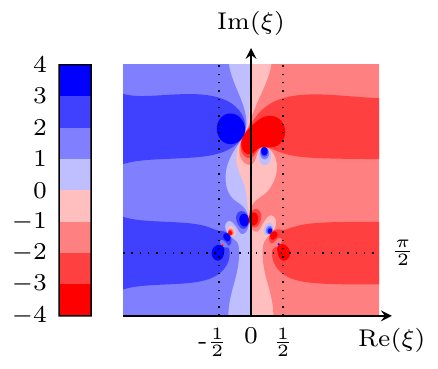}
}
\subfloat[$\mathrm{Im}(\hat{u}_0)$]{
\includegraphics[scale=1.00]{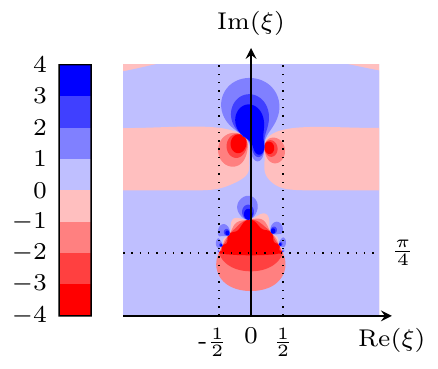}
}
    \caption{(a) A comparison between the AAA approximated function $\hat{u}_0$ and the exact leading-order solution $u_0$ for $\kappa = 1$. The analytic continuation of $\hat{u}_0$ is shown in (b)--(c). The analytically-continued approximation contains two dominant singularity pairs, with three additional singularities located between them. Figures (d)--(f) show the same quantities for $\kappa = 2$. Again, the analytically-continued approximation contains two dominant singularity pairs, with three additional singularities located between them.}
    \label{f:singularityAAA_Toda}
\end{figure}

\subsubsection{Exponential Asymptotic Analysis}

We now apply the AAA algorithm to the numerical leading-order solitary wave to obtain an explicit rational approximation $\hat{u}_0$. To do this, we consider the solution behaviour within a finite domain, which we denote as $[-d, d]$, and use this as the input to the AAA algorithm. The spatial resolution, which we denote as $r$, is given by $\Delta t \cdot \xi$. For $\kappa = 1$, we select $d = 4.7004$ and $r = 0.0118$. For $\kappa = 2$, we select $d = 3.6232$ and $r = 0.0181$. The error tolerance of the AAA algorithm represents the $L^{\infty}$ norm of the difference between the input function and the rational approximation -- reducing the size of the error tolerance has the effect of increasing the order of the polynomials in the numerator and denominator of the rational expression, and therefore the number of singularity pairs in the approximation. For the purpose of this analysis, we select an error tolerance of $10^{-13}$.

Figure~\ref{f:singularityAAA_Toda}(a) and (d) show a comparison between the approximated rational function $\hat{u}_0$ and the exact leading-order solution $u_0$ for $\kappa = 1$ and $\kappa = 2$ respectively. The analytic continuation of $\hat{u}_0$ is shown in Figure~\ref{f:singularityAAA_Toda}(b)--(c) and (e)--(f) for $\kappa = 1$ and $\kappa = 2$ respectively. In both cases, there two dominant singularity pairs, with three more singularity pairs spread along a curve between the dominant pairs. This suggests that leading-order solitary wave profiles with a similar shape produce similar analytically-continued behaviour when approximated using the AAA algorithm. This was not true in the tanh-fitted approximation \eqref{f:amplitude_tanh}, in which the analytic continuation changed significantly between $\kappa = 1$ and $\kappa = 2$.

\begin{figure}[tb]
    \centering
    \includegraphics[scale=1.00]{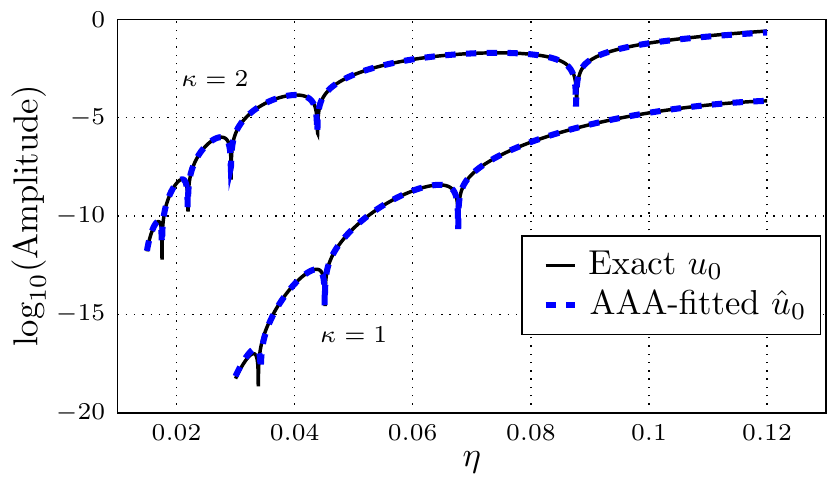}
    \caption{Comparison between the asymptotic amplitude predicted using the AAA-approximation for the leading-order, $\hat{u}_0$, and the exact leading-order behaviour, $u_0$, for $\kappa=1$ and $\kappa=2$. The predicted amplitudes show very strong agreement.}
    \label{f:amplitude_AAA}
\end{figure}

We use the approximated function $\hat{u}_0$ as the leading-order solution in an exponential asymptotic analysis. Each singularity pair generates a set of oscillations with the form given by~\eqref{e:woodpile_sn}. The total oscillatory contribution behind the leading-order wave is given by the superposition of the oscillations generated by each pair. The results of this analysis are presented in in Figure~\ref{f:amplitude_AAA} for $\kappa = 1$ and $2$. These predicted behaviour is visually indistinguishable from the results obtained using $u_0$, and the AAA-approximation method accurately predicts both the amplitude of the oscillations and the values of $\eta$ at which the oscillations vanish.

\begin{figure}
    \centering
\subfloat[Tolerance = $10^{-5}$]{
\includegraphics[scale=1.00]{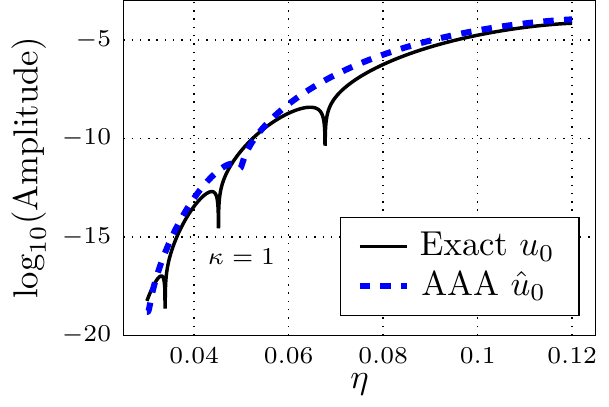}
}
\subfloat[Tolerance = $10^{-9}$]{
\includegraphics[scale=1.00]{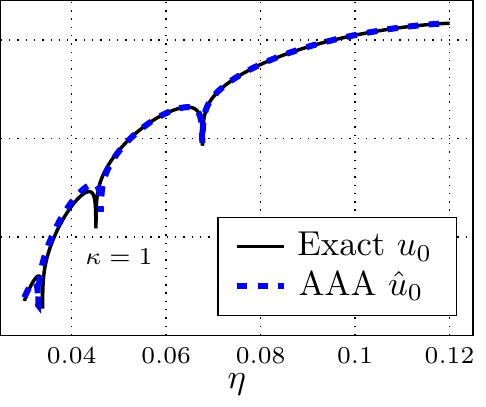}
}
\subfloat[Tolerance = $10^{-13}$]{
\includegraphics[scale=1.00]{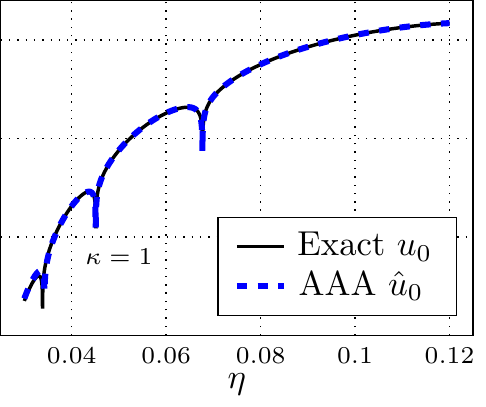}
}
\caption{Comparison of oscillation amplitudes found using the AAA-approximated $\hat{u}_0$ and the exact leading-order $u_0$ with error tolerances in the AAA algorithm of (a) $10^{-5}$, (b) $10^{-9}$ and (c) $10^{-13}$. The relative error $R$ is $0.0695$, $0.0232$ and $0.0104$ for (a), (b) and (c) respectively.}
\label{f:error_Toda}
 \end{figure}

To demonstrate the effect of adjusting the tolerance, we repeated this method with tolerances of $10^{-5}$ and $10^{-9}$. The results of this are shown in Figure~\ref{f:error_Toda}. If the tolerance is relaxed, the approximation becomes less accurate. To quantify the accuracy of each simulation, we introduce an error expression
\begin{align}
R = \frac{\|\log_{10}(\mathrm{Amplitude}_{\mathrm{exact}})-\log_{10}(\mathrm{Amplitude}_{\mathrm{AAA}})\|}
{\|\log_{10}(\mathrm{Amplitude}_{\mathrm{AAA}})\|},
\label{e:errorR}
\end{align}
where $\|\textbf{f}\|^2=\sum_{i=1}^N f_i^2$ and $\textbf{f}=(f_1,f_2,\dots f_N)$. The amplitude vectors contain the oscillation amplitude at 90001 different values of $\eta$, evenly spaced in the domain $[0.03,0.12]$. This expression was chosen as a quantitative measure of the visual fit of the logarithmic data. For AAA error tolerances of $10^{-5}$,  $10^{-9}$ and  $10^{-13}$, the computed values of $R$ are $0.0695$, $0.0232$  and $0.0104$ respectively. This indicates that the fit increases in accuracy as the tolerance is reduced, as would be expected.

\begin{figure}[tb]
\centering
\includegraphics[scale=1.00]{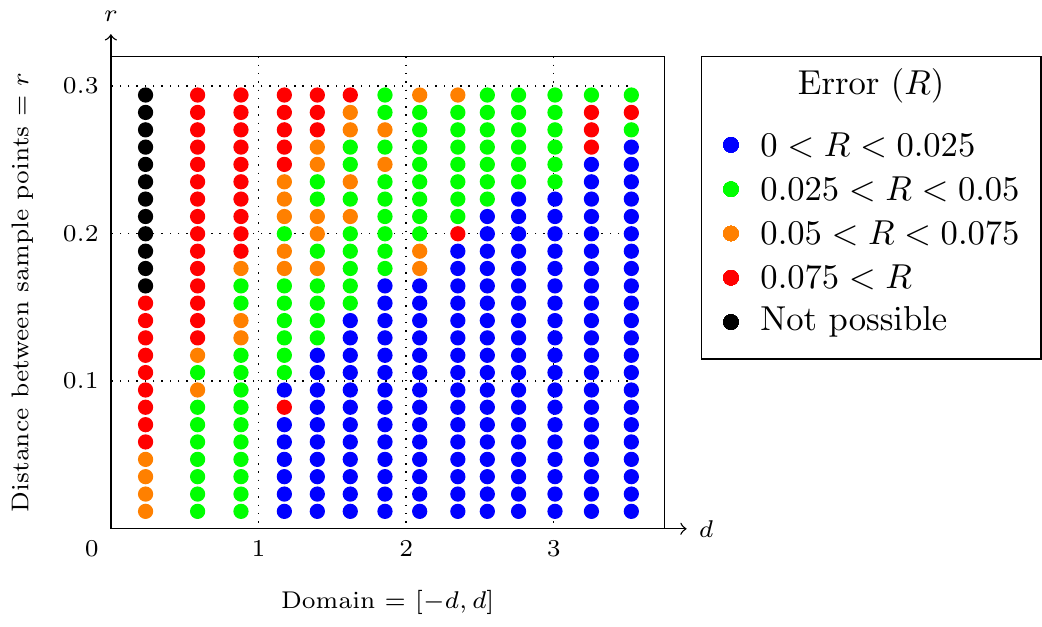}
\caption{The relative error $R$ for the amplitude of the trailing oscillations for different values of the sample domain $[-d, d]$ and the resolution $r$. Increasing the width of the sample domain or decreasing the sample resolution improves the accuracy of the asymptotic behaviour predicted by the exact leading order solution.}
\label{f:Error}
 \end{figure}

To illustrate the effects of adjusting the sample domain and resolution, we repeat this algorithm with different values for the sample resolution $r$ and the domain $d$, and calculate the error $R$ in each case. A summary of these calculations is presented graphically in Figure~\ref{f:Error}. Parameter choices which produce an error of $R < 0.025$ correspond to output which is visibly similar to that given in Figure~\ref{f:error_Toda}(b)--(c), while $R > 0.05$ typically produces output which is similar to that of Figure~\ref{f:error_Toda}(a). Figure~\ref{f:Error} shows that the error can be reduced by increasing the width of the sample domain or decreasing the sample resolution.

\begin{figure}
\centering
\subfloat[Exact: $\mathrm{Re}(u_0)$]{
\includegraphics[scale=1.00]{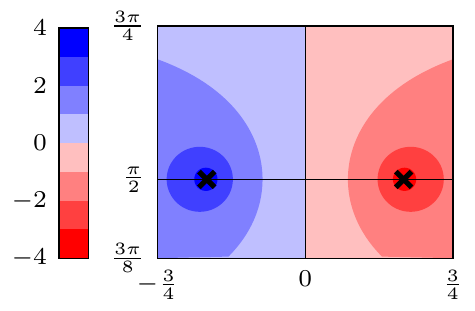}
}
\subfloat[Tanh: $\mathrm{Re}(\hat{u}_0)$]{
\includegraphics[scale=1.00]{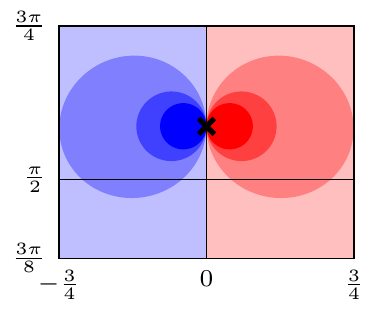}
}
\subfloat[Pad\'{e}: $\mathrm{Re}(\hat{u}_0)$]{
\includegraphics[scale=1.00]{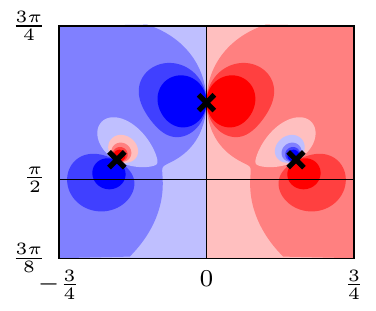}
}
\subfloat[AAA: $\mathrm{Re}(\hat{u}_0)$]{
\includegraphics[scale=1.00]{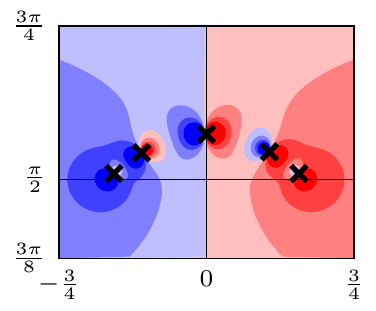}
}

\caption{The locations of singularities in the analytic continuation of the (a) exact solution $u_0$, and (b)--(d) the three different approximated solutions obtained in this section, with $\kappa = 1$ and singular points indicated by crosses. The amplitude of the oscillations is determined by $\mathrm{Im}(\xi_s)$, and the phase is determined by $\mathrm{Re}(\xi_s)$. In the exact leading-order behaviour, the dominant singularity pairs are located in at $\xi = -1/2 \pm \mathrm{i}\pi/2$ and $\xi = 1/2 \pm \mathrm{i}\pi/2$. AAA approximation most accurately predicts the dominant singularity locations, and is therefore able to most accurately reproduce the asymptotic predictions made using the exact solution.}
\label{f:singularities}
\end{figure}

In Figure \ref{f:singularities}, we compare the location of singularities in the analytic continuation of (a) the exact solution, (b) the tanh-fitted approximation, (c) the Pad\'{e} approximant, and (d) the AAA approximation for the leading order behaviour. We see from this image that the AAA approximation most accurately predicts the location of the dominant singularity pairs. The location of the dominant singularity pair strongly influences on both the amplitude and phase of the oscillations, as seen in the asymptotic expression \eqref{e:woodpile_sn}. This is consistent with the observation that the behaviour of the oscillations that was generated using a AAA approximation for the leading order behaviour agrees very strongly with the asymptotic behaviour produced using the exact solutions; furthermore, it outperforms the other tested methods.

Singularities in the exact solutions are logarithmic, while the singularities in each of the approximations for the leading-order behaviour are simple poles. This does not appear to have a significant impact on the woodpile problem, due to the linear coupling between the two governing equations. This will not necessarily be true for more complicated nonlinear problems -- we will discuss this in Section \ref{s:conclusion}. 

\section{Woodpile chain with power-law interaction potential}
\label{s:woodHertz}

In the previous section, we established that the exponentially small oscillations present in nanoptera can be calculated using a AAA approximation in place of the exact leading-order solitary wave. We now apply this method in order to predict behaviour in a system where no exact leading-order solution exists.

The governing equations for woodpile chains with power-law interaction potential \eqref{e:potentialHertz}, with $c = 1/(\alpha+1)$ for simplicity, are given by
\begin{align}\label{1:gov1Hertz}\nonumber
	\ddot{u}(n,t) =  &[\Delta+u(n-1,t)-u(n,t)]_+^\alpha \\&- [\Delta  + u(n,t)-u(n+1,t)]_+^\alpha-k [{u}(n,t) - {v}(n,t)]\,,\\
\eta^2 \ddot{{v}}(n,t) =& {k} [{u}(n,t)-{v}(n,t)]\,.
\label{1:gov2Hertz}
\end{align}
For this section, we consider woodpile chains with zero precompression, as in~\cite{Deng2022}. We therefore set $\Delta=0$. 

In the limit $\eta \rightarrow 0$ we expand $u(x,t)$ and $v(x,t)$ as asymptotic power series in $\eta^2$, as in \eqref{1:asympseriesToda}. We substutitue the series expression~\eqref{1:asympseriesToda} into~\eqref{1:gov2Hertz} match at the leading order in the limit $\eta\to0$, and introduce the co-moving frame $\xi=n-ct$, where $c$ is the solitary wave velocity. We find that $u_0(\xi) = v_0(\xi)$, and
\begin{align}
c^2{u}''_0(\xi) &= [u(\xi-1)-u(\xi,t)]_+^\alpha - [ u(\xi)-u(\xi+1)]_+^\alpha\,,
\label{1:woodpile12Hertz_xi}
\end{align}
where prime denotes derivative with respect to $\xi$. In contrast to woodpile chains with Toda interaction potential, $u_0(\xi)$ cannot be expressed analytically. 

Assuming we have some leading-order solution $u_0$, and we analytically continue the expression so that $\xi \in \mathbb{C}$, we will again find singularity pairs in the complex plane that generate Stokes curves. We label the singularity locations as $\xi = \xi_s$ and $\xi_s^*$, where $\mathrm{Im}(\xi_s) > 0$. We assume that the local behaviour of the solution near the singularity is given by $u_0 \sim \mu (\xi - \xi_s)^{\nu}$ as $\xi \to \xi_s$, for some $\mu$ and $\nu$. A similar exponential asymptotic analysis to Section \ref{s:woodToda} shows that the exponentially small contribution has an identical form to the oscillations in the Toda woodpile chain, presented in \eqref{e:woodpile_sn}.

In~\cite{Deng2022}, the authors approximate the leading-order solution using tanh-fitting, motivated by the work of \cite{sen2001}. Applying an exponential asymptotic analysis allows the amplitude to be accurately predicted for a number of different parameters; however, the asymptotic predictions differ significantly from the numerical simulation for smaller interaction exponents, such as the Hertzian interaction ($\alpha = 1.5$). In this section, we will show that the behaviour of the system can be accurately predicted if we instead use a AAA approximation for the leading-order behaviour.

We obtain the leading-order solitary wave by applying a velocity impulse to a monatomic particle chain as in Section~\ref{ss:Toda_AAA}. Unlike the Toda woodpile chain simulations, we fix the size of the velocity impulse so that $\vec{v}_{61}(0)=5$. We then determine the amplitude of the solitary wave as part of the solution. The simulations were performed with identical parameters to the Toda chain, from Table~\ref{t:parameter}.

\begin{figure}[tb]
\centering
\subfloat[Leading-order $\hat{u}_0(\xi)$ for $\alpha = 1.5$]{
\includegraphics[scale=1.00]{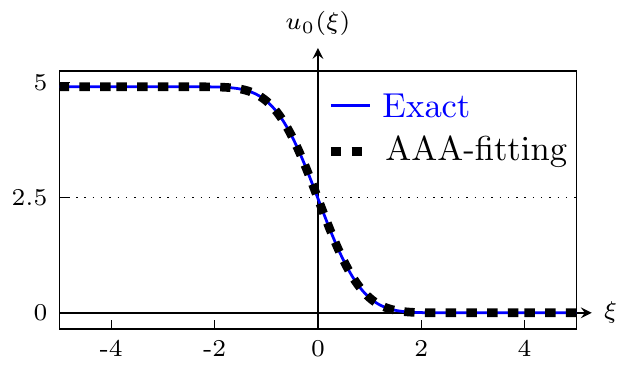}
}
\subfloat[$\mathrm{Re}(\hat{u}_0)$]{
\includegraphics[scale=1.00]{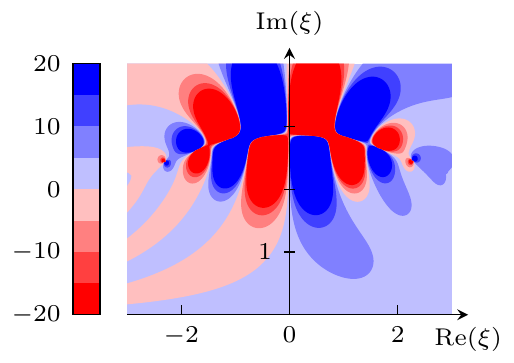}
}
\subfloat[$\mathrm{Im}(\hat{u}_0)$]{
\includegraphics[scale=1.00]{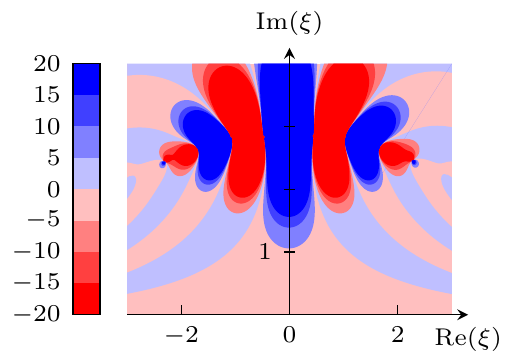}
}
    \caption{(a) A comparison between the AAA approximated function $\hat{u}_0$ and the exact leading-order solution $u_0$ for $\kappa = 1$. The analytic continuation of $\hat{u}_0$ is shown in (b)--(c). }
    \label{f:displacement_compare}
\end{figure}

We then use the AAA algorithm to obtain a rational approximation for the leading-order behaviour, which we denote $\hat{u}_0$. For the purpose of this study, we use the parameters $d = 4.6296$ and $r = 0.0231$, and the AAA tolerance is $10^{-13}$. In Figure~\ref{f:displacement_compare}(a), we compare $\hat{u}_0$ to the numerical solution. 

\begin{figure}[tb]
\centering
\subfloat[Singularities: $\alpha = 1.5$]{
\includegraphics[scale=1.00]{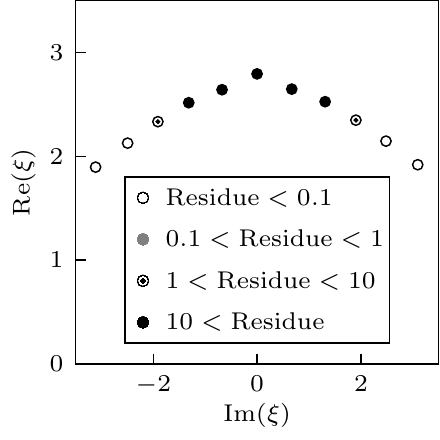}
}
\subfloat[Singularities: $\alpha = 1.75$]{
\includegraphics[scale=1.00]{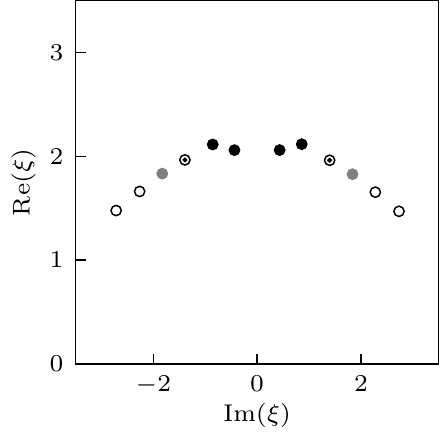}
}
\subfloat[Singularities: $\alpha = 2$]{
\includegraphics[scale=1.00]{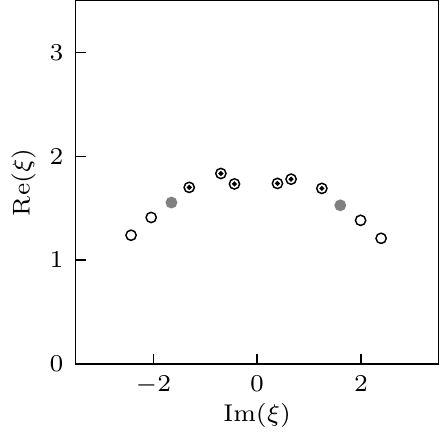}
}

\subfloat[Singularities: $\alpha = 2.5$]{
\includegraphics[scale=1.00]{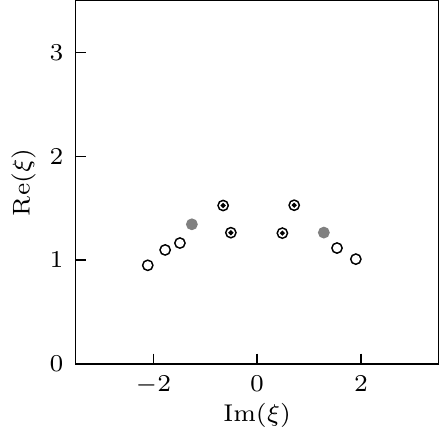}
}
\subfloat[Singularities: $\alpha = 3$]{
\includegraphics[scale=1.00]{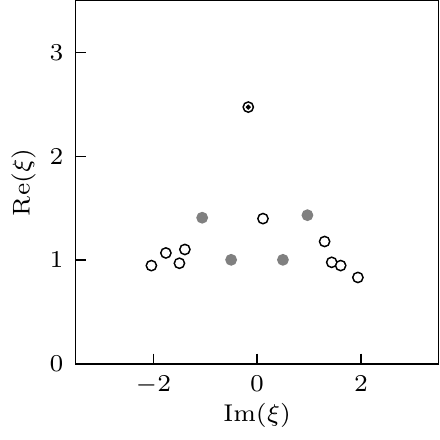}
}
\subfloat[Singularities: $\alpha = 3.5$]{
\includegraphics[scale=1.00]{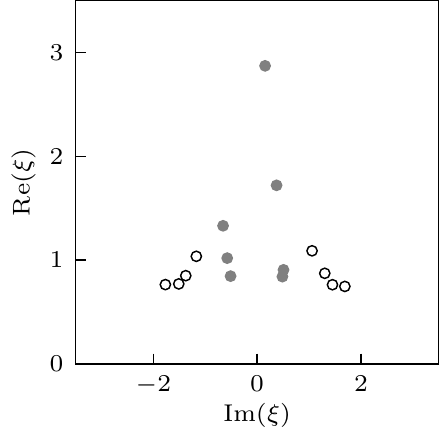}
}
    \caption{Singularities with positive imaginary part for $\alpha=1.5,1.75,2,2.5,3,3.5$. }
    \label{f:singularityAAA_Hertz}
\end{figure}

We analytically continue $\hat{u}_0$, and determine the location of the singularites in the complex plane. These singularities are all simple poles, and are shown in Figure~\ref{f:singularityAAA_Hertz}. We also illustrate the size of the residual associated with each singularity. In each case, there are pairs which are further from the real axis than the dominant singularity pair, but which have much larger residuals. This means that they can still have a measurable effect on the oscillatory behaviour. 

This complicated pole structure explains why the tanh-fitting method was unable to accurately capture details of the oscillation behaviour. The tanh-fitting method did not accurately predict singularity locations that are further from the real axis than the dominant singularity pair, and these singularities have a significant affect on the solution behaviour due to their large residuals.

\begin{figure}[tb]
    \centering
    \subfloat[$\alpha = 1.5$]{
    \includegraphics[scale=1.00]{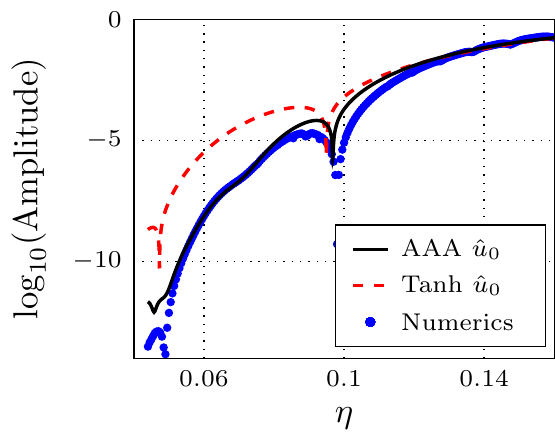}
    }
        \subfloat[$\alpha = 1.75$]{
    \includegraphics[scale=1.00]{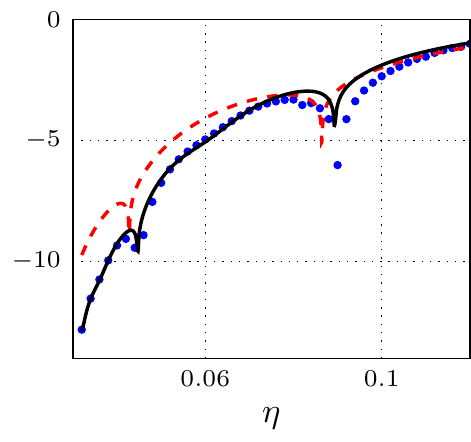}
    }
        \subfloat[$\alpha = 2$]{
    \includegraphics[scale=1.00]{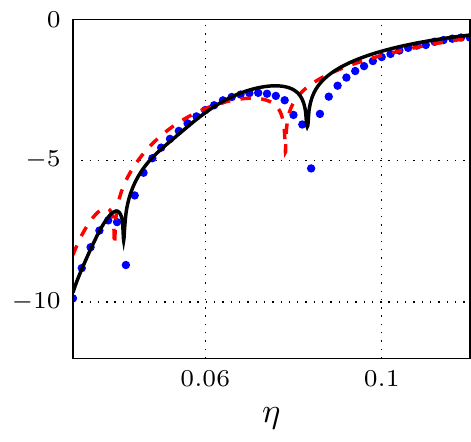}
    }
    
        \subfloat[$\alpha = 2.5$]{
    \includegraphics[scale=1.00]{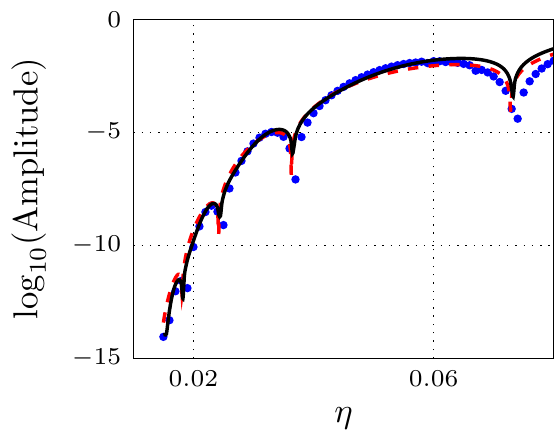}
    }
        \subfloat[$\alpha = 3.5$]{
    \includegraphics[scale=1.00]{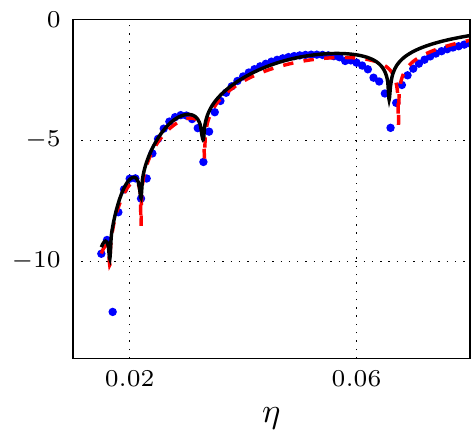}
    }
        \subfloat[$\alpha = 3.5$]{
    \includegraphics[scale=1.00]{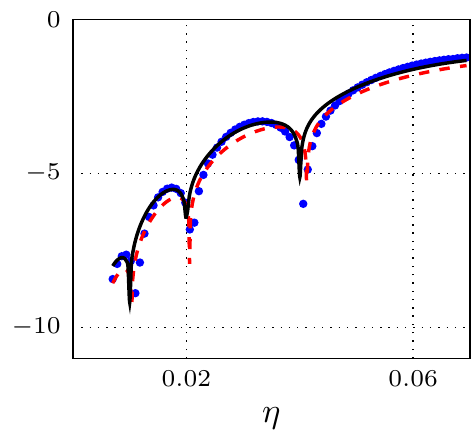}
    }
    \caption{Comparison of the oscillation amplitudes predicted by asymptotics based on AAA-approximated leading-order solutions (black), based on leading order with tanh-fitted leading-order solutions (red, dashed) and based on numerical simulations (blue dots). For larger values of $\alpha$, both the AAA and tanh-fitted asymptotic predictions are accurate. For smaller values of $\alpha$,  predictions made based on AAA-approximation are more accurate than tanh-fitting. The AAA-approximation method can accurately describe small variations in the amplitude due to interactions between subdominant contributions. This is particularly apparent at $\eta \approx 0.075$ for $\alpha = 1.5$.}
    \label{f:amplitude_AAA_Hertz}
\end{figure}

To determine the behaviour of the exponentially small oscillations, we take the sum of the contributions from every singularity pair, and then determine the amplitude of this combined expressions. The contributions for each singularity pair are given by \eqref{e:woodpile_sn}. In Figure~\ref{f:amplitude_AAA_Hertz}, we present the results for a range of interaction exponents $\alpha$, and compare these predictions to those made in \cite{Deng2022} using the tanh-fitted method. For larger values of $\alpha$ both methods accurately predict the system behaviour, but for smaller values of $\alpha$, the predictions made using AAA approximation are significantly more accurate than those made using tanh-fitting.

Using AAA approximation to generate $\hat{u}_0$ also allows the asymptotic behaviour that occurs due to subdominant poles. This can be seen most clearly in Figure~\ref{f:amplitude_AAA_Hertz}(a), which shows the results for $\alpha = 1.5$. In addition to accurately predicting the values of $\eta$ at which cancellation occurs, our results also correctly describe the smaller variations in the amplitude that can be seen between these values. These smaller variations occur due to the cancellation of subdominant oscillatory contributions, generated by singularity pairs that are further from the real axis than the dominant pairs.

\section{Conclusion}
\label{s:conclusion}

We applied three approximation methods to study nanoptera in Toda woodpile chains -- tanh-fitting, Pad\'{e} approximants, and the AAA approximation algorithm -- and compared the results with asymptotic predictions made using the exact leading-order solution. We found that, while all three methods were capable of describing aspects of the exponentially small behaviour, the AAA approximation was most effective in determining two key features of the exponentially small oscillations in a Toda woodpile chains: the amplitude of the oscillations, and the values of the mass ratio at which the oscillations vanish.

Using the AAA algorithm to approximate the leading-order behaviour has several advantages over tanh-fitting and Pad\'{e} approximants. Approximations made using tanh-fitting cannot be made arbitrarily precise, as we cannot take series coefficients indefinitely. Pad\'{e} approximants rely entirely on the behaviour of the numerical solution at a point, which requires taking progressively higher numerical derivatives to increase the accuracy of the rational approximation. AAA approximation avoids this by fitting the solution over a specified domain, rather than a single point. AAA approximation made the most accurate predictions of the singularity locations, explaining its accurate prediction of exponentially small behaviour.

We used the AAA algorithm to approximate the leading-order behaviour of woodpile chains with power-law interaction. The subsequent asymptotic predictions agree with numerics well for all values of the interaction exponent $\alpha$. This is notable, as~\cite{Deng2022} studied this problem using tanh-fitting methods, and found that the method did not accurately describe the oscillations for moderate or small values of the interaction exponent, including Hertzian interactions ($\alpha = 1.5$).

For $\alpha=1.5$ our asymptotic prediction also captures details in the asymptotic behaviour that were missed by the tanh-fitting method. By looking at the pole locations, it is apparent why this occured. This method was able to identify smaller variations in the amplitude that are caused by interactions between subdominant poles that are not accurately identified using tanh-fitting -- in some cases, such as $\alpha = 1.5$, these poles have large residues, and therefore can visibly affect the solution behaviour. Asymptotic behaviour generated by this method can include significant effects, even if they are exponentially subdominant compared to the oscillation amplitude.

The next step to establishing the viability of this method is to apply the AAA approximation to determine the leading-order behaviour for more complicated model systems such as diatomic particle chains~\cite{Lustri, Lustri1}, which includes additional challenges.The AAA approximation is effective at locating singularities, which it represents as simple poles. In the case of branch points, the approximation contains a cluster of simple poles that accurately approximate the branch cut behaviour. The order of the singularity plays a significant role in nonlinear problems, and it is likely that the local approximation of $u_0$ near the singularity will need to be determined. Data from~\cite{Trefethen2021} indicates that the manner in which the poles cluster near the singular point depends on the order of the singularity. If the order of the singularity can be determined from this clustering, the appropriate branch cut could be included in the approximation function with the weight determined through the least-squares process.

\begin{figure}
\centering
\subfloat[$\log_{10}|\mathrm{Re}(u_0 - \hat{u}_0)|$]{
\includegraphics[scale=1.00]{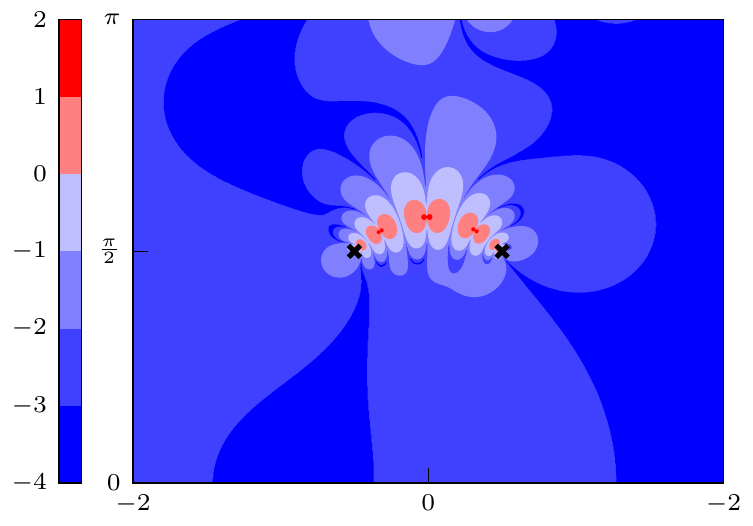}
}
\subfloat[$\log_{10}|\mathrm{Im}(u_0 - \hat{u}_0)|$]{
\includegraphics[scale=1.00]{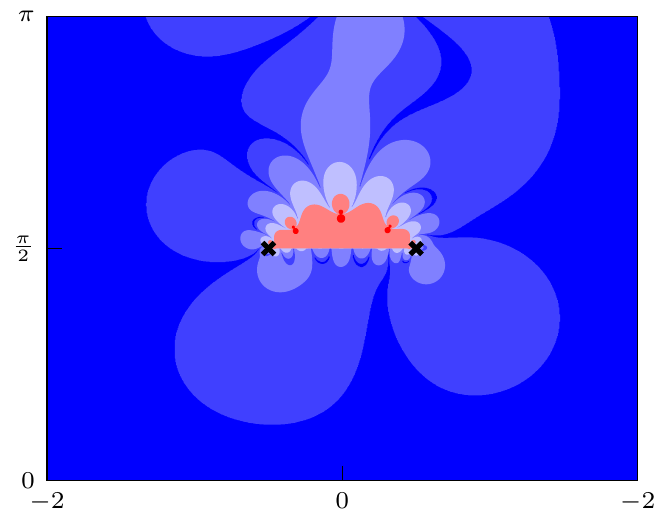}
}
\caption{Error of the AAA approximant $\hat{u}_0$ of the leading-order behaviour $u_0$ for the Toda lattice with $\kappa = 1$. The poles of the exact solution are shown as black crosses. The AAA approximant closely approximates the exact solution except in a region near the poles. }
\label{f:conclusion}
\end{figure}

Evaluating the singulant for nonlinearly coupled chains will require the evaluation of complex contour integrals of $u_0$. Figure~\ref{f:conclusion} shows the difference between the AAA approximation for the Toda chain and the exact leading-order behaviour, which is small except in a region near the poles. The singularity positions must be considered in prescribing an integral contour, to control the accuracy of the approximation.

\appendix
\section{Detailed exponential asymptotic analysis}\label{A:ExpAsymp}
The method described in this section is very similar to that of~\cite{Lustri,Lustri1,Deng2021,Deng2022}, and we therefore omit much of the intermediate analysis.

\subsection{Late-order terms}
In the limit $\eta \rightarrow 0$ we write $u(x,t)$ and $v(x,t)$ as asymptotic power series in $\eta^2$, giving the series expressions from \eqref{1:asympseriesToda}. Writing the governing equations \eqref{1:gov1Toda}--\eqref{1:gov2Toda} in terms of $\xi$ and matching at each order of $\eta$ gives
\begin{align}\label{2:govser1}
	\nonumber& c^2 u_j''(\xi) = (u_j(\xi-1) - u_j(\xi))
    \exp{(u_0(\xi-1) - u_0(\xi))} \\&\quad- (u_j(\xi) - u_j(\xi+1))\exp{(u_0(\xi) - u_0(\xi+1))} - k [u_j(\xi) - v_j(\xi)] + \ldots\,,\\
	&c^2 v_{j-1}''(\xi) = k [u_j(\xi) - v_j(\xi)]\,.
	\label{2:govser2}
\end{align}
We only retain terms containing $u_j$, $v_j$, and derivatives of $u_j$ and $v_{j-1}$, as these terms govern the late-order asymptotic behaviour. 

We assume that $u_0$ contains a singularity at a point in the complex plane $\xi = \xi_s$. We consider two cases: the case where the singularity in $u_0$ is logarithmic, so that $u_0 \sim \mu \log(\xi - \xi_s)$ as $\xi \to \xi_s$, and the case the singularity has asymptotic behaviour $u_0\sim \mu(\xi - \xi_s)^{\nu}$ as $\xi \to \xi_s$. In the case of the exact solution \eqref{e:LO_precise}, $\mu = 1$ for the singularities at $\xi = \xi_{1,\pm,j}$ and $\mu = -1$ for the singularities at $\xi = \xi_{2,\pm,j}$.

We pose a late-order ansatz that consists of sums of terms with the form
\begin{equation}\label{2:ansatz}
	u_j \sim \frac{U(\xi)\Gamma(rj + \beta_1)}{\chi(\xi)^{rj + \beta_1}}\,, \quad v_j \sim \frac{V(\xi)\Gamma(rj + \beta_2)}{\chi(\xi)^{rj + \beta_2}} \quad \mathrm{as} \quad j \rightarrow \infty\,,
\end{equation}
where $r$ is the number of times that $u_{j-1}$ and $v_{j-1}$ must be differentiated to obtain $u_j$ and $v_j$, as illustrated in Section~\ref{s:exponential}. From~\eqref{2:govser1}--\eqref{2:govser2} we have $r=2$.
We set $\chi=0$ at $\xi = \xi_s$ so that late-order terms are singular at the same location as the leading-order solution.
By substituting the late-order ansatz~\eqref{2:ansatz} into~\eqref{2:govser1}, we find that only $\beta_1 + 2 = \beta_2$ can produce a nontrivial asymptotic balance. This implies that $u_j = \mathcal{O}(v_{j-1})$ as $j \rightarrow \infty$, and therefore that $v_j \gg u_j$ as $j\to\infty$.

Applying the late-order ansatz \eqref{2:ansatz} to~\eqref{2:govser2} gives
\begin{align}\nonumber
	 \frac{c^2 (\chi'(\xi))^2 V(\xi)\Gamma(2j + \beta_2)}{\chi(\xi)^{2j + \beta_2}} - \frac{2c^2 \chi'(\xi) V'(\xi)\Gamma(2j + \beta_2-1)}{\chi(\xi)^{2j + \beta_2-1}}&\\
 	- \frac{c^2 \chi''(\xi) V(\xi) \Gamma(2j + \beta_2 - 1)}{\chi(\xi)^{2j + \beta_2 - 1}} + \cdots =& -\frac{k V(\xi)\Gamma(2j + \beta_2)}{\chi(\xi)^{2j + \beta_2}} + \cdots\,,
\end{align}
where the omitted terms are $\mathcal{O}(v_{j-1})$ in the $j \rightarrow \infty$ limit. Matching terms at $\mathcal{O}(v_j)$ in the $j \rightarrow \infty$ limit, we obtain $c^2(\chi'(\xi))^2 = -k$, implying $\chi'(\xi) = \pm \mathrm{i}\sqrt{k}/c$, and therefore
\begin{equation}\label{2:singulant}
	\chi(\xi) = \pm\tfrac{ \mathrm{i}}{c} \sqrt{k}(\xi-\xi_s)\,.
\end{equation}
Stokes curves occur where $\mathrm{Im}(\chi) = 0$ and $\mathrm{Re}(\chi) > 0$~\cite{BerryHowls1990}. This corresponds to the positive sign choice for singularities in the upper half plane, and the negative sign choice for those in the lower half plane. We do not consider late-order terms associated with the remaining sign choices, as they do not generate Stokes curves.

Matching terms at $\mathcal{O}(v'_{j-1})$, we obtain the prefactor equation $2 V'(\xi)\chi'(\xi) = 0$; $V$ is therefore constant. The strength of the singularities must be consistent with the leading-order behaviour in the limit that $\xi \to \xi_s$. For these expressions to be consistent, we require $\beta_2 = \nu$ and $\beta_1 = \nu + 2$.

We follow a procedure similar to that of~\cite{Deng2021,Deng2022}, and determine $V$ by matching the outer expansion for the late-order terms \eqref{2:ansatz} with an inner solution in the neighbourhood of $\xi_s$. Through this analysis, we find that $V = \mu$ if the singularity is logarithmic, and $V =\mathrm{i}  \mu \sqrt{k}/c$ if the singularity is a pole or a branch point.

\subsection{Stokes switching}
Truncating the series~\eqref{1:asympseriesToda} after $N$ terms gives
\begin{equation}\label{3:series}
	u(\xi) = \sum_{j=0}^{N-1} \eta^{2j} u_j(\xi) + S_N(\xi)\,,  \quad  v(\xi) = \sum_{j=0}^{N-1} \eta^{2j} v_j(\xi) + R_N(\xi)\,,
\end{equation}
where $S_N$ and $R_N$ are the remainder terms obtained by truncating the series, and are exponentially small if we optimally truncate the series. We denote the optimal truncation point as $N = N_{\mathrm{opt}}$. The heuristic in~\cite{Boyd1999} shows that $N_{\mathrm{opt}}$ is determined by finding the point where consecutive terms in the series are equal in size, giving $N_{\mathrm{opt}}=|\chi|/2\eta+\omega$, where $\omega\in[0,1)$ is chosen to ensure that $N_{\mathrm{opt}}$ is an integer. 

Inserting~\eqref{3:series} into~\eqref{1:gov1Toda}--\eqref{1:gov2Toda} gives governing equations for $R_N$ and $S_N$ as $\eta\to 0$:
\begin{align}\nonumber
&c^2 S''_N(\xi) \sim   \,[S_N(\xi + 1) - S(\xi)]\e^{-[u_0(\xi+1) - u_0(\xi)]} \\&\quad\quad  - [S_N(\xi) - S_N(\xi-1)]\e^{-[u_0(\xi) - u_0(\xi-1)]} - k[S_N(\xi) - R_N(\xi)] + \ldots,\\
&\eta^2 c^2 R''(\xi) \sim  \, k [S_N(\xi) - R_N(\xi)] - \eta^{2N} v''_{N-1}(\xi),
\end{align}
where omitted terms are asymptotically subdominant as $\eta\to 0$. We also find from a balancing argument that $R_N(\xi) \gg S_N(\xi)$ in this limit, and we may therefore simplify this expression -- in particular, the second equation decouples from the first. Noting that $N_{\mathrm{opt}}$ is large as $\eta \to 0$, we apply \eqref{2:ansatz} for the late-order terms, giving
\begin{equation} 
\eta^2 c^2 R''(\xi) + k \chi'(\xi)^2 R_N(\xi) \sim -\eta^{2N}\frac{V \Gamma(2N + \beta_2)}{\chi(\xi)^{2N + \beta_2}} \qquad \mathrm{as} \qquad \eta \to 0.
\end{equation}
The right-hand side is asymptotically subdominant compared to the balance on the left-hand side except for a narrow region in the neighbourhood of the Stokes curve. We use a WKB analysis to show that the remainder can be written as $R_N \propto  \e^{-\chi/\eta}$ as $\eta\to 0$ away from the Stokes curve. To capture the behaviour of the solution near the Stokes curve, we write the ansatz $R_N = A(\xi) \e^{-\chi/\eta}$ as $\eta\to 0$. By writing $\chi = r \e^{\mathrm{i}\theta}$, the variation of $A$ in terms of $\theta$ is given by
\begin{equation}
\diff{A}{\theta} \sim \frac{V \chi' \Gamma(2N + \beta_2)}{2\chi^{2N + \beta_2}} \e^{\chi/\eta} \qquad \mathrm{as} \qquad \eta \to 0.
\end{equation}
The Stokes curve follows $\theta = 0$. We can define an inner region near the Stokes curve in terms of an inner variable $\hat{\theta} = \eta^{-1/2}\theta$, and use asymptotic matching to determine the variation of $A$ in this region. We find that $A = 0$ on the left-hand side of the Stokes curve, and $A \sim \mathrm{i} \pi V/\eta^{\nu}$ as $\xi\to 0$ on the right-hand side of the Stokes curve. 

The contribution given by the singularity at $\xi = \xi_s^*$ is the complex conjugate of the contribution given by the singularity at $\xi = \xi_s$. Hence, the exponentially small behaviour that appears on the left-hand side of the Stokes curve connecting the singularity pair $\xi_s$ and $\xi_s^*$ is 
\begin{equation}
	R_N \sim\frac{2|V|   \pi}{\eta^{\nu}}
\exp\left(-\frac{\sqrt{k}\,\mathrm{Im}(\xi_s)}{c\eta}\right)
\cos\left(\frac{\sqrt{k}(\xi-\mathrm{Re}(\xi_s))}{c\eta}-\phi\right)\quad \mathrm{as} \quad \eta \rightarrow 0\,,
\label{e:woodpile_sn_appendix}
\end{equation}
where $\nu = 0$ if the singularity is logarithmic. The quantity $\phi$ is given by $\pi/2$ if the singularity is logarithmic, and the phase of $V$ if it is a pole or branch cut.

\section*{Acknowledgments}
The authors acknowledge the support of Australian Research Council Discovery Project \#190101190.

\bibliography{reference2}
\bibliographystyle{plain}

\end{document}